\documentclass[pra,aps,twocolumn,amsmath,amssymb,floatfix,reprint,footinbib,superscriptaddress,longbibliography]{revtex4-2}
\usepackage{graphics}      
\usepackage{graphicx}      
\usepackage{longtable}     
\usepackage{url}           
\usepackage{bm}    
\usepackage{braket}        
\usepackage{xcolor}
\usepackage{comment}
\usepackage{siunitx}
\usepackage{physics}
\usepackage[T1]{fontenc}
\usepackage{booktabs}
\usepackage{appendix}
\usepackage{amsmath,amssymb,epsfig,color}
\usepackage{mathrsfs}
\usepackage{amsfonts}
\usepackage{dcolumn}
\usepackage{subfigure}
\usepackage{soul}
\usepackage{tabularx}
\usepackage{mleftright} 
\usepackage[colorlinks,linkcolor=blue,citecolor=blue,hyperindex,bookmarks=false,pdfstartview=FitH]{hyperref}
\usepackage[left]{lineno}  
\usepackage{float}
\usepackage{ulem}

\usepackage[T1]{fontenc}
\usepackage[utf8]{inputenc}

\usepackage[margin=2.5cm]{geometry}
\usepackage{setspace}
\newcommand{\beq}{\begin{equation}}
\newcommand{\eeq}{\end{equation}}
\newcommand{\beqa}{\begin{eqnarray}}
\newcommand{\eeqa}{\end{eqnarray}}

\newcommand{\figref}[1]{\mbox{Fig.~\ref{#1}}}
\newcommand{\tabref}[1]{\mbox{Table~\ref{#1}}}
\newcommand{\secref}[1]{\mbox{Sec.~\ref{#1}}}

\renewcommand{\eqref}[1]{\mbox{Eq.~(\ref{#1})}}

\newcommand{\figpanel}[2]{Fig.~\hyperref[#1]{\ref*{#1}(#2)}}
\newcommand{\figpanels}[3]{Fig.~\hyperref[#1]{\ref*{#1}(#2)--(#3)}}
\newcommand{\figpanelNoPrefix}[2]{\hyperref[#1]{\ref*{#1}(#2)}}

\usepackage{etoolbox}

\makeatletter
\newcommand{\SMtableofcontents}{%
  \begin{center}
    \textbf{\large Contents}
  \end{center}
  \@starttoc{smtoc}%
}
\makeatother

\begin{document}
\title{Optimal robust single-qubit gates in superconducting quantum processors}

\author{Tangyou Huang}

\author{Anuj Aggarwal}

\author{Jingjun Zhu}
\email{jingjun@chalmers.se}

\author{Akshay Gaikwad}

\author{David Niepce}
\affiliation{Department of Microtechnology and Nanoscience, Chalmers University of Technology, 41296 Gothenburg, Sweden}
 
\author{Marcus Rommel}
\author{Andreas Nylander}
\author{Amr Osman}
\author{Janka Biznárová}
\author{Anita Fadavi Roudsari}
\affiliation{Department of Microtechnology and Nanoscience, Chalmers University of Technology, 41296 Gothenburg, Sweden}

\author{Stéphane Guérin}
\affiliation{Université Bourgogne Europe, CNRS, Laboratoire Interdisciplinaire Carnot de Bourgogne ICB, UMR 6303, 21000 Dijon, France}

\author{Gheorghe Sorin Paraoanu}
\affiliation{Department of Applied Physics and InstituteQ, Aalto University, FI-00076 Aalto, Finland}

\author{Anton Frisk Kockum}

\author{Giovanna Tancredi}
\affiliation{Department of Microtechnology and Nanoscience, Chalmers University of Technology, 41296 Gothenburg, Sweden}


\date{\today}

\begin{abstract}

Fault-tolerant quantum computing requires high-fidelity quantum gates that are resilient to control errors and device imperfections. Here, we develop energy-optimal control pulses for single-qubit gates that are simultaneously robust against frequency detuning and waveform inhomogeneities while also suppressing leakage outside the computational subspace. Combining the optimized pulse envelopes with an adaptive control scheme, we experimentally demonstrate optimal robust gates on a superconducting quantum processor. For a gate duration of \SI{90}{\nano\second}, we achieve fidelities above \SI{99.9}{\percent} for the $X$ gate and \SI{99.6}{\percent} for Hadamard-type gates. For the $X$ gate, we maintain fidelities above \SI{99.5}{\percent} for simultaneous drive-amplitude variations up to \SI{20}{\percent} and qubit detuning up to \SI{4}{\mega\hertz}. This robustness outperforms that of conventional Gaussian pulses by an order of magnitude, without much difference in peak gate fidelity.
These results establish a route towards robust single-qubit control in large-scale quantum processors, advancing the development of fault-tolerant quantum computing.

\end{abstract}

\maketitle

 
\onecolumngrid

\section{Introduction}
Realizing fault-tolerant quantum computation through quantum error correction demands single-qubit gates that remain reliable despite control imperfections inherent to physical hardware~\cite{Dennis2002TopologicalMemory, Bravyi2024HighThresholdLDPC, Paetznick2026ImprovedLogicalErrorRates, Park2025TransducedNoise, Bluvstein2026FaultTolerantNeutralAtomArchitecture, GoogleQuantumAI2025BelowThreshold}. In superconducting processors, imperfect microwave pulses and flux crosstalk can cause coherent errors and shift qubit frequencies~\cite{Rol2020Cryoscopepulse, Hutchings2017frequencyshift},
while the finite anharmonicity of qubits permits unwanted excitation outside the computational subspace~\cite{DRAG}.
These effects both degrade gate fidelity and increase calibration overhead, motivating pulse-design strategies that simultaneously address drive-amplitude errors, frequency detuning, and leakage suppression.
While periodic automated calibration and reinforcement-learning methods can mitigate temporal qubit frequency drift after it occurs~\cite{sivak2026reinforcement, sivak2023real},
error-robust gate design offers a complementary route: suppressing sensitivity to dominant errors at the pulse-design level~\cite{Preskill2018NISQ, Kandala2019ErrorMitigation}.

In recent years, instead of error-dependent numerical schemes~\cite{tf39-n6s3, universalRobust2024prl, robust_DRL,4zzl-qbsf}, perturbative frameworks have enabled the development of robust-control schemes within two-level systems~\cite{composite2011prl, STArobust2012optimally, RobustSSSP, NumRobust2015prl, PhysRevA.109.012616, wang2012composite} that preserve gate performance against either amplitude noise or frequency detuning.
More recent progress based on geometrical optimization~\cite{zeng2019geometric, Geometrical2021prxq, universalRobust2024prl, Dapeng2024prl, Geometric2025prA} has sought to achieve simultaneous resilience to both error sources.
However, these constructions rely on a pre-specified ansatz and therefore cannot guarantee time- or energy-optimal implementations---a concern of growing importance given the ongoing debate over the energy costs of quantum technologies~\cite{Auffeves1, Auffeves2} and the challenges of cooling large superconducting processors~\cite{Mohseni2026}.

The framework of robust inverse optimal (RIO) pulses~\cite{RIO2020prl, RIOdigital2023prl} addresses these shortcomings by producing optimal gate operations while maintaining robustness against both qubit-frequency detuning and control-pulse envelope distortions~\cite{steph2026pra}.
However, RIO-based pulses are obtained by inverse engineering of the pulse envelope from optimized state trajectories in a two-level system, which usually leads to abrupt temporal variations~\cite{RIO2020prl, RIOdigital2023prl, steph2026pra}, such as square-pulse-like features.
These sharp features are strongly distorted by realistic control hardware with finite bandwidth, leading to leakage errors in systems with finite anharmonicity~\cite{DRAG}, thereby degrading experimental performance and leaving RIO pulses largely unexplored experimentally.

In this article, we introduce time-rescaling robust inverse-optimal (TRIO) pulses by combining the RIO framework with an adaptive pulse-engineering technique.
We experimentally demonstrate TRIO pulses for single-qubit gate operations using superconducting transmon qubits.
For the $X$ gate, we achieve a gate fidelity exceeding \SI{99.9}{\percent} with a \SI{90}{\nano\second} TRIO pulse. This pulse maintains a substantially higher fidelity ($\sim$\SI{99.5}{\percent}) than a conventional Gaussian pulse ($\sim$\SI{92}{\percent}) of the same length under combined errors of up to \SI{4}{\mega\hertz} in frequency detuning and \SI{20}{\percent} in waveform amplitude mismatch; the Gaussian pulse only tolerates an order of magnitude smaller errors to maintain the same fidelity, and displays slightly better peak fidelity.
We further verify the feasibility and robustness of \SI{90}{\nano\second} Hadamard-type single-qubit gates ($X_{\pi/2}$ and $Y_{-\pi/2}$) with fidelities above \SI{99.6}{\percent} using the TRIO framework, indicating that the scheme can be extended toward a universal set of single-qubit gates.
Taken together, our results provide the first experimental demonstration of optimal robust gates and establish a practical foundation for deploying such robust control schemes as alternatives to conventional pulses in large-scale quantum processors.

\begin{figure*} 
\centering
\includegraphics[width=0.65\columnwidth]{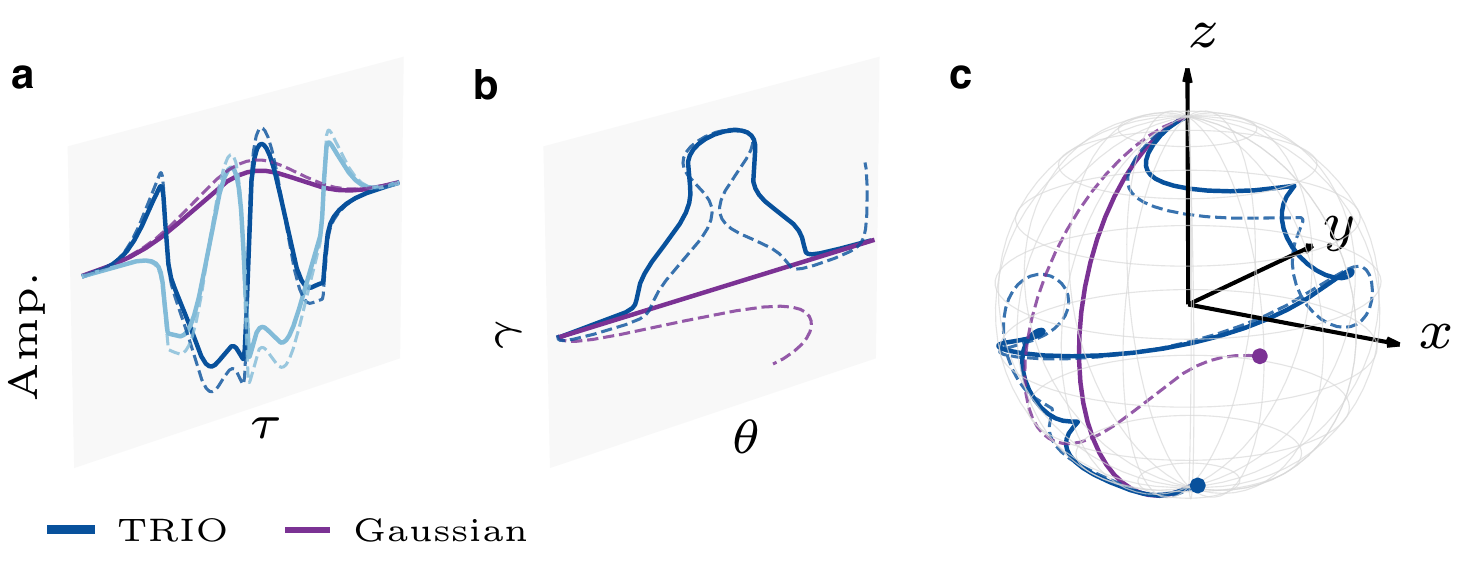}
\caption{\textbf{The TRIO pulse.}
Comparison of TRIO (blue) and Gaussian (purple) pulses for an $X$ gate in a two-level system.
\textbf{a}, Pulse envelopes; dark and light blue denote the real (dark blue) and imaginary (light blue) components of the TRIO pulse, respectively.
\textbf{b}, Corresponding trajectories in the $(\theta,\gamma)$ plane, where $\theta$ and $\gamma$ denote the mixing angle and global phase of the evolving state, respectively.
\textbf{c}, State trajectories on the Bloch sphere.
Solid curves represent ideal control, whereas dashed curves include a simultaneous $\SI{20}{\percent}$ drive-amplitude offset and a $\SI{4}{\mega\hertz}$ detuning error.
Filled circles mark the final states of the dashed trajectories.
\label{fig0:schematic}}
\end{figure*}


\section{Results}

We implement TRIO pulses on a single qubit in a superconducting processor containing two fixed-frequency transmon qubits~\cite{Koch2007} coupled through a flux-tunable coupler. Device parameters and experimental setup are reported in the Supplemental Material~\cite{SM}. The TRIO  pulses are implemented through in-phase and quadrature (IQ) control, together with a compensating virtual-$Z$ rotation $V_Z(\phi)$, whose phase is determined from the optimal-control solution and refined experimentally~\cite{SM}.
To benchmark robustness, we introduce a controlled qubit-frequency detuning by adjusting the coupler's DC flux bias while we apply amplitude errors by directly modifying the output amplitude of the pulse generator.
Gate fidelity and leakage are characterized using interleaved randomized benchmarking (IRB)~\cite{Magesan2012IRB}. Throughout this work, we compare TRIO pulses with conventional Gaussian pulses with an envelope
$\Omega_{\mathrm{G}}(t) = A \exp[-(t-t_f/2)^2/(2\sigma^2)]$,
where $A$ is the pulse amplitude, $t_f$ is the gate duration, and $\sigma = t_f/6$ unless otherwise specified.

Figure~\ref{fig0:schematic} compares the TRIO and Gaussian pulse envelopes and their corresponding trajectories for $X$ gate in a two-level system.
The TRIO envelope is smooth at the boundaries, compatible with experimental implementation, while admitting the {designed trajectory} geometry in the $(\theta, \gamma)$ representation~\cite{steph2026pra}, where $\theta$ and $\gamma$ denote the qubit mixing angle and the global phase, respectively. The dashed curves in \figpanels{fig0:schematic}{a}{c} show the perturbed evolution in the presence of an amplitude error $\alpha=\SI{20}{\percent}$ and a detuning error $\delta=\SI{4}{\mega\hertz}$.
Although the perturbed TRIO trajectory deviates from the ideal path during the evolution, it reconverges near the target state, whereas the Gaussian trajectory exhibits a substantial final-state deviation.
Consequently, numerical simulations yield gate errors of approximately $10^{-4}$ for TRIO pulses and $10^{-1}$ for Gaussian pulses.
This comparison indicates that the enhanced robustness of the TRIO pulse arises from the reduced sensitivity of its geometric trajectory to coherent control errors~\cite{steph2026pra, RIO2020prl}. Unlike a resonant Gaussian $\pi$-pulse, whose fixed rotation axis causes systematic amplitude and detuning errors to accumulate, TRIO simultaneously modulates the Rabi frequency and detuning to continuously modify the rotation axis on the Bloch sphere. These engineered rotations allow error contributions generated at later stage of the evolution to compensate those accumulated earlier, thereby enhancing robustness while preserving the desired target operation. 
Details of the TRIO pulse construction are provided in the Methods and in the Supplemental Material~\cite{SM}.


\begin{figure*} 
\centering
\includegraphics[width=1\columnwidth]{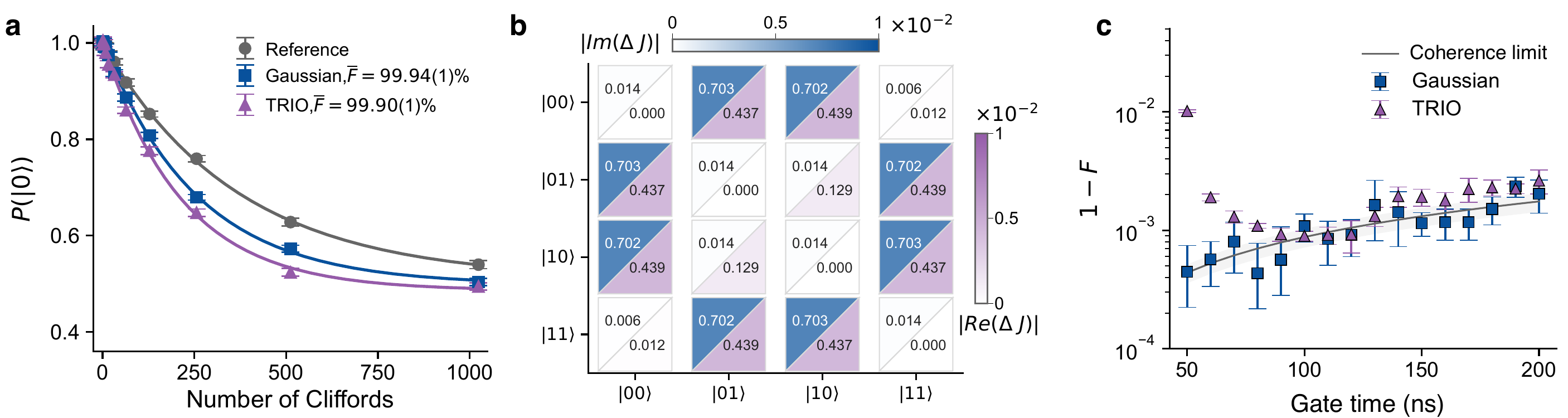}
\caption{\textbf{Experimental characterization of the ${X}$ gate.}
\textbf{a}, Reference and interleaved randomized benchmarking decay curves, yielding gate fidelities of \SI{99.94(1)}{\percent} for the Gaussian pulse and \SI{99.90(1)}{\percent} for the TRIO pulse, averaged over five experiments.
\textbf{b}, Absolute values of the real and imaginary components of $\Delta J$ for the TRIO-based X gate, i.e., the difference between the experimentally reconstructed and ideal $X$-gate Choi matrices.
\textbf{c}, Gate infidelity as a function of gate duration for the Gaussian and TRIO pulses, compared with the coherence limit~\cite{Abad2022} (solid curve).
}
\label{fig:figure1}
\end{figure*}

Figure~\ref{fig:figure1} presents the benchmarking of the TRIO-based $X$ gate and its Gaussian counterpart for a gate duration of $t_f = \SI{90}{\nano\second}$.
All reported gate fidelities for both Gaussian and TRIO pulses include calibrated DRAG (derivative removal by adiabatic gate) corrections~\cite{DRAG}.
IRB measurements averaged over five independent experiments yield $X$-gate fidelities of $\overline{F} = 99.90 \pm \SI{0.01}{\percent}$ for the TRIO pulse and $99.94 \pm \SI{0.01}{\percent}$ for the Gaussian pulse, as shown in \figpanel{fig:figure1}{a}. 

To further characterize the TRIO-based ${X}$ gate, we perform quantum process tomography (QPT)~\cite{Chuang1997QPT} and reconstruct its Choi matrix~\cite{hashim-prxquantum-2025}, which provides a complete representation of the quantum process.
Figure~\figpanelNoPrefix{fig:figure1}{b} shows the absolute values of the real and imaginary components of the difference 
$\Delta J = J_{\rm exp}-J_{\rm ideal}$, 
between the experimentally reconstructed Choi matrix $J_{\rm exp}$ and the ideal Choi matrix $J_{\rm ideal}$.
The small deviations indicate close agreement between the implemented and target operations, with an average estimated gate fidelity of $99.93 \pm \SI{0.05}{\percent}$ over ten independent QPT experiments.

We further perform IRB over gate durations ranging from 50 to \SI{200}{\nano\second}, as shown in \figpanel{fig:figure1}{c}. For gate times $t_f < \SI{90}{\nano\second}$, the larger area of the TRIO pulse ($3.093\pi$; cf.~\figref{fig0:schematic}) requires a higher-amplitude envelope, leading to increased leakage $(\sim 10^{-3})$. This is significantly higher than the $(\sim 10^{-5})$ leakage of the Gaussian-based $\pi$ pulse~\cite{SM} and results in the observed lower gate fidelity at short time scales.

\begin{figure*} 
\centering
\includegraphics[width=1\columnwidth]{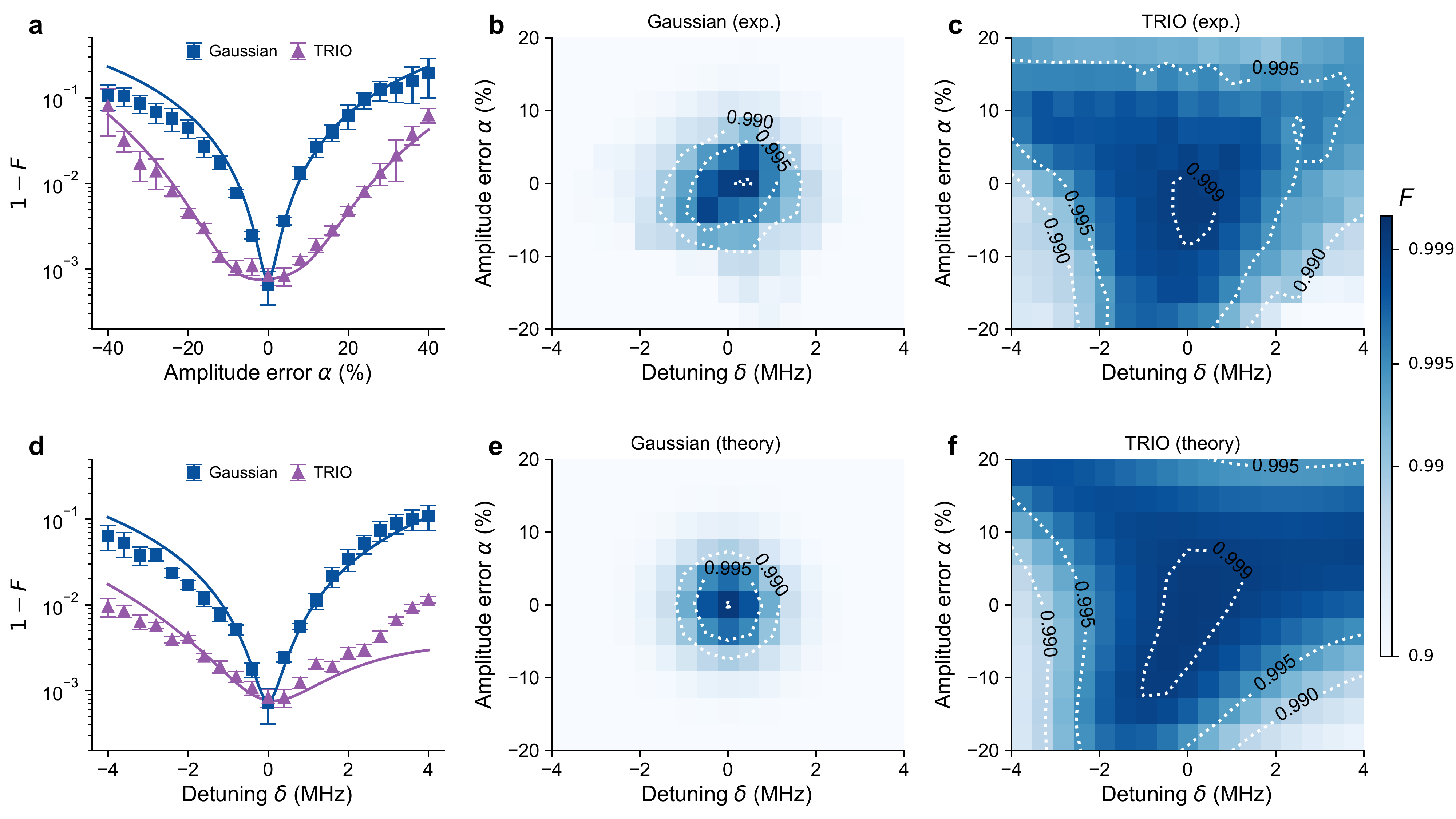}
\caption{\textbf{Robustness against amplitude and detuning errors for a $t_f = \SI{90}{\nano\second}$ $X$ gate.}
\textbf{a}, \textbf{d}, IRB infidelity for the Gaussian and TRIO pulses as a function of relative amplitude error $\alpha$ and frequency detuning $\delta$, respectively.
Data points are experimental results, with error bars denoting the standard deviation over ten repetitions. Solid curves show numerical simulations.
\textbf{b}, \textbf{c}, Experimentally measured two-dimensional gate-fidelity maps under simultaneous amplitude and detuning errors for the Gaussian and TRIO pulses, respectively. 
\textbf{e}, \textbf{f}, Numerical simulations corresponding to the cases in \textbf{b} and \textbf{c}, respectively.
Dotted contours indicate the labeled fidelity levels.
}
\label{fig:robust_exp}
\end{figure*}

In \figref{fig:robust_exp}, we demonstrate the robustness of the TRIO pulse in comparison to the standard Gaussian pulse for both amplitude errors $\alpha$ and frequency-detuning errors $\delta$. Figure~\figpanelNoPrefix{fig:robust_exp}{a} and \figpanel{fig:robust_exp}{d} illustrate the measured infidelity (data points) for both pulses as a function of $\alpha$ and $\delta$, respectively. These data demonstrate that the Gaussian pulse exhibits a sharp infidelity minimum around the error-free point ($\alpha=0$, $\delta=0$) for each error type, whereas the TRIO pulse sacrifices a small amount of peak fidelity to flatten the infidelity response over a much wider error interval, thus outperforming the Gaussian pulse in terms of robustness.
The solid curves show numerical simulations that account for decoherence, using the measured average relaxation time $T_1 = \SI{75.4}{\micro\second}$ and dephasing time $T_2^* = \SI{55.1}{\micro\second}$. 

In \figpanel{fig:robust_exp}{b} and \figpanel{fig:robust_exp}{c}, we show the measured infidelity of the Gaussian and TRIO pulse, respectively, as a function of both $\alpha$ and $\delta$. For the Gaussian pulse, the high-fidelity region is narrow and tilted, indicating strong sensitivity to combined amplitude and detuning offsets. For the TRIO pulse, the contours enclosing regions with $\mathcal{F} \geq \SI{99}{\percent}$ and $\mathcal{F} \geq \SI{99.5}{\percent}$ extend over a substantially larger fraction of the tested parameter space (an order of magnitude larger in both $\alpha$ and $\delta$), showing that the same pulse remains robust when both error sources are present simultaneously. Infidelity simulations are reported in \figpanel{fig:robust_exp}{e} and \figpanel{fig:robust_exp}{f} for the Gaussian and TRIO pulse, respectively, showing very good agreement with data.

We next extend the TRIO framework beyond the $X$ gate. Following the RIO construction in Ref.~\cite{steph2026pra}, we obtain an energy-optimal pulse with a pulse area of $3.395\pi$ that implements the Hadamard ($H$)-type operation 
$U_{\mathrm{H};\varphi} = \frac{1}{\sqrt{2}} 
\mleft(
\begin{smallmatrix}
1 & -e^{i\varphi}\\
e^{-i\varphi} & 1
\end{smallmatrix}
\mright)$,
where $\varphi=0.73\pi$.
Two virtual-$Z$ rotations convert this operation into an $H$ gate, up to a global phase:
$H_{\phi_1, \phi_2}^{\mathrm{TRIO}}
= V_Z(\phi_1) U_{\mathrm{H};\varphi} V_Z(\phi_2)$, 
with $\phi_1 = \varphi$ and $\phi_2 = \pi-\varphi$,
where operators act from right to left. After experimental phase calibration, the TRIO-based $H$ gate achieves an IRB fidelity of approximately \SI{99.63(1)}{\percent}.
For comparison, a Gaussian-based implementation, 
$H^{\mathrm{G}}=V_Z(\pi)Y_{-\pi/2}^{\mathrm{G}}$
up to a global phase, achieves a fidelity of \SI{99.92(1)}{\percent}.

\begin{figure*}
\centering
\includegraphics[width=0.65\columnwidth]{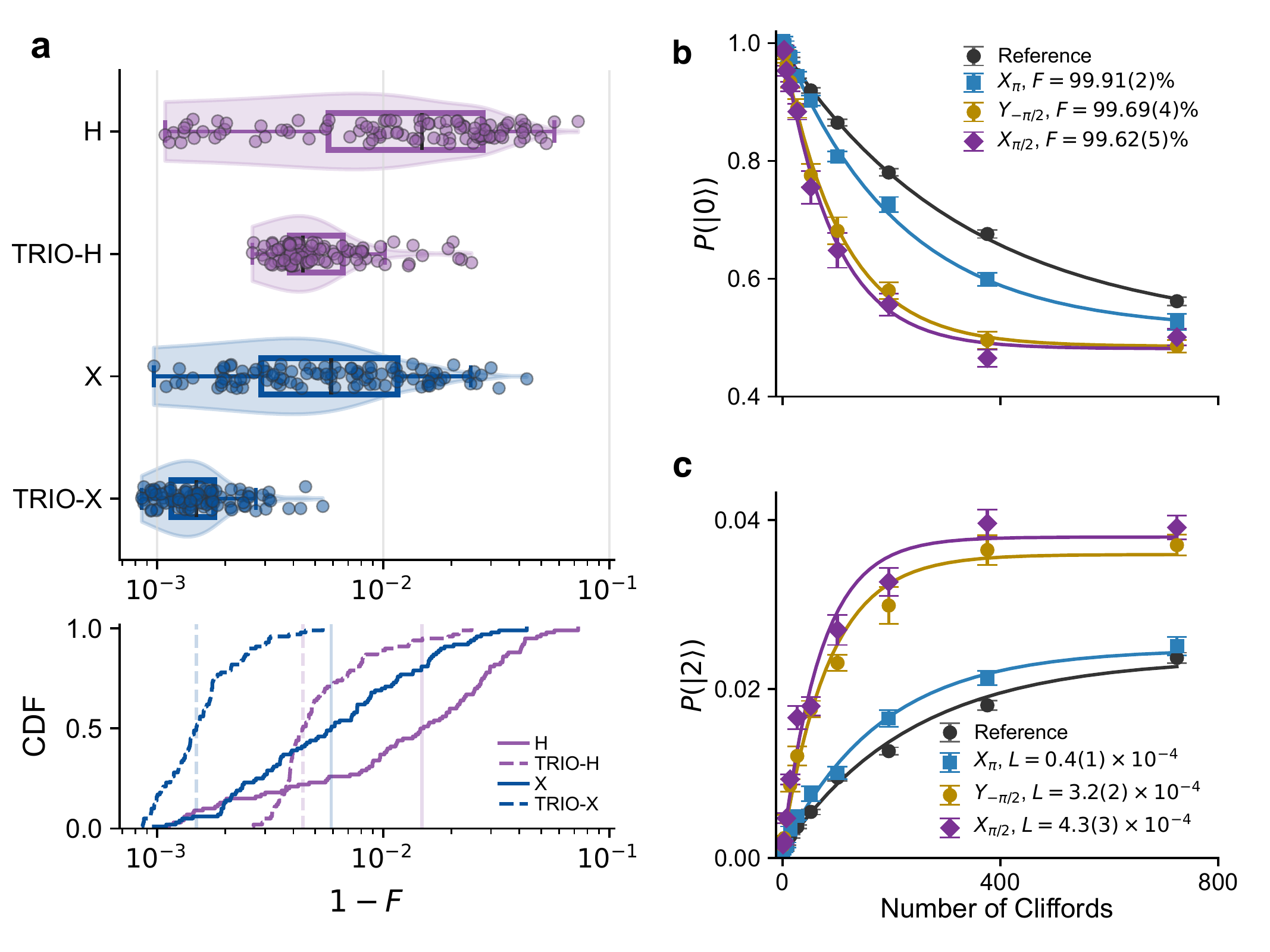}
\caption{\textbf{Universal gate set and robustness statistics.}
\textbf{a}, Statistical distributions of gate infidelity for the $H$ and $X$ gates and their TRIO counterparts under randomly sampled drive-amplitude and frequency-detuning errors, with $\alpha \in [-20,20]\,\%$ and $\delta \in [-2,2]\,\mathrm{MHz}$ drawn from truncated normal distributions. The upper panel shows the distribution of infidelities. The boxes indicate the interquartile range with the central lines marking the medians. The lower panel shows the corresponding cumulative distribution functions, with vertical lines marking the mean values.
\textbf{b-c}, Survival probability $P(|0\rangle)$ and leaked population $P(|2\rangle)$ as a function of the number of Clifford gates for the fidelity and leakage IRB, respectively. The IRB curves of the gate set $\{ {X}, {Y}_{-\pi/2},{X}_{\pi/2}\}$ are compared to the reference curve using the Gaussian-based pulse.  
}
\label{fig:gate-statistics}
\end{figure*}

Adjusting the virtual-$Z$ rotations also yields
$X_{\pi/2} = H_{0.23\pi,-0.23\pi}^{\mathrm{TRIO}}$
and
$Y_{-\pi/2}=H_{-0.27\pi,0.27\pi}^{\mathrm{TRIO}}$.
Together with $X$, these operations form a universal single-qubit gate set, whose experimental characterization is shown in \figpanel{fig:gate-statistics}{b}.
Interleaved randomized benchmarking yields fidelities of \SI{99.91(2)}{\percent}, \SI{99.69(4)}{\percent} and \SI{99.62(5)}{\percent} for
$X$, $Y_{-\pi/2}$, and $X_{\pi/2}$, respectively.
In \figpanel{fig:gate-statistics}{c}, the two $\pi/2$-type gates exhibit greater leakage than $X$, consistent with the larger pulse area required by the underlying $H$-type construction at the same gate duration.

To assess the robustness of this universal gate set, we independently sample the relative amplitude error and frequency detuning from zero-mean Gaussian distributions,
$\alpha\sim\mathcal{N}(0,\sigma_\alpha^2)$ and $\delta\sim\mathcal{N}(0,\sigma_\delta^2)$, 
with $\sigma_\alpha=0.2/3$ and $\sigma_\delta=(2/3)\,\mathrm{MHz}$. 
The corresponding three-standard-deviation intervals are 
$\alpha \in [-20,20]\,\%$ and $\delta \in [-2,2]\,\mathrm{MHz}$.
For each error realization, the same amplitude and detuning offsets are applied to the Gaussian and TRIO pulses. Repeating this procedure over independent realizations yields the infidelity distributions shown in \figpanel{fig:gate-statistics}{a}. For both the $X$ and $H$-type gates, the TRIO pulses exhibit lower mean infidelities and distributions concentrated at lower infidelity values than their Gaussian counterparts. The corresponding empirical cumulative distribution functions [lower panel in \figpanel{fig:gate-statistics}{a}]  
show that TRIO pulses shift the gate-infidelity distributions towards lower values for both gate types.


\section{Conclusion and discussion}

We have experimentally demonstrated time-rescaling robust inverse-optimal (TRIO) control for a universal single-qubit gate set on a superconducting processor.
By combining robust, energy-optimal trajectories with a time-rescaling technique, we obtain smooth, hardware-compatible waveforms that retain robustness against simultaneous amplitude and detuning errors while suppressing leakage outside the computational subspace.
Our results establish a practical basis for robust single-qubit gates, with direct applications to scalable superconducting quantum computation.

Compared with conventional Gaussian-based pulses, TRIO pulses exhibit 
{high fidelity with} robustness over a broader range of control parameters {at an optimal energy}.
This feature is particularly relevant for large-scale processors, where calibration drift, crosstalk, and device variations can compromise gate performance.
To pursue shorter gate durations, further improvements could combine the present approach with higher-bandwidth waveform generation and active leakage mitigation~\cite{PRXQuantum.5.030353} or {high-order $N$-level DRAG~\cite{PRL-KAMDRAG}}.
The framework may also benefit from implementation in systems with larger anharmonicities or longer coherence times, including fluxonium~\cite{Fluxonium1, Fluxonium2}.
Our numerical simulations predict TRIO gate errors below $10^{-6}$ for an anharmonicity of \SI{1}{\giga\hertz} (see Appendix C in the Supplemental Material~\cite{SM}).
Broadly, the TRIO techniques can be implemented in other quantum-computing architectures~\cite{Park2025TransducedNoise, Bluvstein2026FaultTolerantNeutralAtomArchitecture, GoogleQuantumAI2025BelowThreshold} and have applications in quantum technologies such as quantum sensing~\cite{gt51-g34c,chang2026benchmarkingdarkmattersearch}, for which precise control is often challenged by control imperfections and environmental noise.


\section{Methods}

\subsection{Robust optimal control}

We consider a time-dependent driven two-level system described by the Hamiltonian
\begin{equation}
H_{\alpha,\delta}(t) = H_{0}(t) + \lambda V(t)
= \frac{\hbar}{2} \mleft[ \Omega'(t)\sigma_x - \Delta'(t)\sigma_z \mright] ,
\end{equation}
where
$H_{0}(t) = \hbar \mleft[ \Omega(t)\sigma_x - \Delta(t)\sigma_z \mright] /2$
is the ideal Hamiltonian and $\lambda V(t)$ are quasi-static control imperfections parameterized by $\lambda \equiv (\alpha, \delta)$. Here, $\alpha$ and $\delta$ denote the relative amplitude miscalibration and detuning offset, respectively, giving rise to the effective control fields $\Omega'(t) = (1 + \alpha)\Omega(t)$ and $\Delta'(t) = \Delta(t) + \delta$. The ideal dynamics is parameterized by a smooth Bloch-sphere trajectory,
\begin{equation}
\ket{\phi_0(t)} = e^{-i \gamma(t)/2} \mleft[ e^{i \varphi(t)/2} \cos \mleft( \theta(t)/2 \mright), e^{-i \varphi(t)/2} \sin \mleft( \theta(t)/2 \mright) \mright] ,
\end{equation}
from which the control fields are reconstructed through inverse trajectory engineering,
\begin{align}
\Delta(t) &= \dot{\varphi}(t) - \dot{\gamma}(t) \cos\theta(t) , \\
\Omega(t) &= \sqrt{\dot{\theta}^2(t) + \dot{\gamma}^{2}(t) \sin^2 \theta(t)} ,
\end{align}
where the mixing angle 
$\theta(t) \equiv \theta \in [0, \pi]$, 
the relative phase 
$\varphi(t) \equiv \varphi \in [-\pi, \pi]$,
and the global phase 
$\gamma(t) \equiv \gamma \in [0,2\pi]$.

Within the perturbative geometric framework developed in Refs.~\cite{SM, RIO2020prl}, robust gate design is reduced to identifying trajectories $(\theta(t), \gamma(t))$ that implement the target gate while suppressing the leading perturbative effects induced by $\lambda V(t)$ (see Supplemental Material~\cite{SM}). Energy-optimal robust trajectories $(\theta^{\rm opt}(t), \gamma^{\rm opt}(t))$ are obtained by minimizing the pulse energy, i.e., $\mathcal{E} = \int \Omega^2 (t) dt$,  subject to the gate and robustness constraints, yielding the optimized control fields $\{\Omega^{\rm opt}(t), \Delta^{\rm opt}(t)\}$ through inverse reconstruction. The complete optimization procedure is presented in the Supplemental Material~\cite{SM}. For the $X$ and $H$-type gates, the resulting energy-optimal pulses have $\mathcal{E}^{\rm opt}_{\rm X} = 3.461 \pi$ and $\mathcal{E}^{\rm opt}_{\rm H}=5.633\pi$, respectively, with pulse areas of $3.093\pi$ and $3.395\pi$, respectively. The corresponding pulse envelopes and Bloch-sphere trajectories are presented in the Supplemental Material~\cite{SM}.


\subsection{The TRIO pulse}

The energy-optimal inverse optimization protocol yields robust control fields that simultaneously suppress quasi-static amplitude and detuning errors while minimizing the pulse energy. However, the resulting waveforms generally exhibit abrupt temporal variations~\cite{RIO2020prl, RIOdigital2023prl, SM}, including square-pulse-like features that are incompatible with finite-bandwidth microwave hardware and can increase leakage errors in weakly anharmonic superconducting qubits. To bridge the gap between the ideal two-level optimization and experimental implementation, we introduce the robust time-rescaling inverse optimization (TRIO) protocol.

The central idea of TRIO is to preserve the energy-optimal geometric trajectory while reconstructing smooth and experimentally feasible control fields in terms of a time-rescaling (TR) scheme~\cite{zhu2022jpb,time-rescaling}. Introducing the rescaled time $\tau$ satisfying
$t=g(\tau)$, with $g(0)=0$ and $g(\tau_f)=t_f$, where $\tau_f$ denotes the final time in the rescaled frame and $t_f$ is the total gate duration,  
the Schr\"{o}dinger evolution remains invariant under the transformation
$H_{\alpha,\delta}(\tau)=g'(\tau)H_{\alpha,\delta}[g(\tau)]$ 
with
$g'(\tau)={dt}/{d\tau}$, 
such that the quantum evolution and the geometric trajectory remain unchanged.

{The above property allows one to replace the optimized Rabi frequency}
$\Omega^{\rm opt}(t)$ {by an experimentally realizable smooth wave-form
$\Omega_{\rm TR}(\tau)$}
and determine the corresponding time-rescaling function
$g(\tau)$
by matching it to the optimal pulse.
The associated detuning profile
$\Delta_{\rm TR}(\tau)$ 
is then reconstructed from the invariant trajectory.
Consequently,
\beq 
\{\Omega^{\rm opt}(t), \Omega_{\rm TR}(\tau)\} \to g(\tau) \to \Delta_{\rm TR}(\tau) \to \{\theta^{\rm opt},\gamma^{\rm opt}\} ,
\eeq
which exactly reproduces the original trajectory in the $(\theta, \gamma)$ space. 

In this work, we choose the rescaled Rabi frequency as a normalized hyper-Gaussian envelope,
\beq
\Omega_{\rm TR}(\tau) = A \exp \mleft[- \mleft( \frac{\tau - t_f/2}{2\sigma} \mright)^4 \mright],
\eeq
where $\int_{0}^{t_f} \Omega_{\rm TR}^2 (\tau) d\tau \approx \mathcal{E}^{\rm opt}$ is approximately conserved, such that the reconstructed pulse inherits the energy scale of the original optimization while providing smooth pulse boundaries. The corresponding time-rescaling function and reconstructed detuning are obtained accordingly, as described in the Supplemental Material~\cite{SM}.

Moreover, the smooth TRIO pulse allows us to apply the standard DRAG correction~\cite{DRAG} to suppresses leakage to higher energy levels while compensating for the AC Stark shift induced by the finite anharmonicity. Retaining the leading-order correction in $1/\xi$, where $\xi$ is the transmon anharmonicity, the resulting control fields, incorporating both leakage suppression and Stark-shift compensation, are given by
\begin{align}
\tilde{\Omega}_{\rm TR}(\tau)
&=
\Omega_{\rm TR}(\tau) - \frac{\Omega_{\rm TR}(\tau)\Delta_{\rm TR}(\tau)}{\xi}
- i\frac{\dot{\Omega}_{\rm TR}(\tau)}{\xi} , \\
\tilde{\Delta}_{\rm TR}(\tau)
&= \Delta_{\rm TR}(\tau) - \frac{\Omega_{\rm TR}^{2}(\tau)}{2\xi} .
\end{align}
We refer to the Supplemental Material~\cite{SM} for the detailed derivation of these results.

To experimentally implement the reconstructed time-dependent detuning profile on a fixed-frequency transmon, we employ microwave phase modulation. The accumulated phase associated with the reconstructed detuning is given by
$\phi_{\rm eff}(\tau) = \int_{0}^{\tau} \Delta_{\rm TR}(s)\,ds$,
which is encoded into the microwave drive through the phase transformation
$W(\tau) = e^{-i\phi_{\rm eff}(\tau)\sigma_z/2}$ 
with the resonant driving.
This transformation converts the reconstructed detuning into a time-dependent drive phase while keeping the carrier frequency fixed, enabling arbitrary detuning profiles to be synthesized experimentally. Consequently, the laboratory-frame microwave envelope of the DRAG-corrected TRIO pulse is
%
\beq
\tilde{\Omega}_{\rm TRIO}(\tau)
= e^{i\int_{0}^{\tau} \tilde{\Delta}_{\rm TR}(s) ds} \tilde{\Omega}_{\rm TR}(\tau)
= \tilde{\Omega}_{x}(\tau) + i\tilde{\Omega}_{y}(\tau) ,
\label{eq:TRIO}
\eeq
where the initial conditions satisfy
$\phi_{\rm eff}(0) = 0$
and 
$\tilde{\Delta}_{\rm TR}(0) = 0$.
In this vein, the complex-valued TRIO pulse can be implemented across diverse quantum hardware platforms by encoding its real and imaginary components into two phase-orthogonal control quadratures, such as the \(I\) and \(Q\) microwave channels of a superconducting quantum processor~\cite{Lucero2010}. 
The same principle applies to the amplitude and phase quadratures of optical or Raman fields in atomic and trapped-ion systems~\cite{Bluvstein2022}.

\section*{Data availability}
The experimental and numerical data supporting this study are available upon reasonable request
\section*{Code availability}
The code for numerical simulation supporting this study are available upon reasonable request
%
\begin{acknowledgments}
We acknowledge useful discussions with Ilya Moskalenko, Marko Kuzmanovi\'c, and Yu-Han Chang from the Aalto team.
The Chalmers team acknowledges financial support by the Knut and Alice Wallenberg through the Wallenberg Center for Quantum Technology (WACQT) and the Horizon Europe programme HORIZON-CL4-2022-QUANTUM-01-SGA via the project 101113946 OpenSuperQPlus100. Device fabrication was performed at Myfab Chalmers. S.G. acknowledges support from the TeLeMAQ LabCom (ANR-25-LCV2-0010-01), the EUR-EIPHI Graduate School (17-EURE-0002), the Conseil Régional de Bourgogne-Franche-Comté, and Dijon Métropole. G.S.P.~acknowledges support from the Research Council of Finland, project no.~376371. A.F.K.~is also supported by the Swedish Foundation for Strategic Research (grant numbers FFL21-0279 and FUS21-0063) and the Norwegian Research Council through the Norwegian Quantum Software Center (NorQSoft, project number 361350).
\end{acknowledgments}
%
%
\section*{Author contributions.} 
{T.H.~and J.Z.~conceived the idea. T.H.~and A.A.~performed the experiments and analyzed the data with assistance from D.N., A.G., and J.Z. ~J.Z.~led the theoretical work with contributions from T.H., S.G., and G.S.P. ~T.H.~and J.Z.~wrote the manuscript, and A.F.K.~and G.T. contributed~to its revision. A.F.K.~and G.T.~supervised the project and provided guidance throughout. A.A., A.N, M.R., A.O., A.F.R. and J.B.~participated in device fabrication. All authors contributed to discussions and the interpretation of the results. }
\section*{Competing interests}
The authors declare no competing interests.
\bibliography{ref}	

@misc{SM,
  note = {Supplemental Material},
}

@article{nielsen2002simple,
  title={A simple formula for the average gate fidelity of a quantum dynamical operation},
  author={Nielsen, Michael A},
  journal={Physics Letters A},
  volume={303},
  number={4},
  pages={249--252},
  year={2002},
  url={https://www.sciencedirect.com/science/article/pii/S0375960102012720},
  publisher={Elsevier}
}

@article{robust_DRL,
  title = {{Experimental Deep Reinforcement Learning for Error-Robust Gate-Set Design on a Superconducting Quantum Computer}},
  author = {Baum, Yuval and Amico, Mirko and Howell, Sean and Hush, Michael and Liuzzi, Maggie and Mundada, Pranav and Merkh, Thomas and Carvalho, Andre R.R. and Biercuk, Michael J.},
  journal = {PRX Quantum},
  volume = {2},
  pages = {040324},
  year = {2021},
  doi = {10.1103/PRXQuantum.2.040324}
}

@article{universalRobust2024prl,
  title = {{Universally Robust Quantum Control}},
  author = {Poggi, Pablo M. and De Chiara, Gabriele and Campbell, Steve and Kiely, Anthony},
  journal = {Physical Review Letters},
  volume = {132},
  pages = {193801},
  year = {2024},
  doi = {10.1103/PhysRevLett.132.193801}
}

@article{tf39-n6s3,
  title = {Superconducting-qubit gates robust to parameter fluctuations},
  author = {Wright, E.M. and Van Damme, L. and Glaser, N.J. and Devra, A. and Roy, F.A. and Englhardt, J. and Bruckmoser, N. and Koch, L. and Marx, A. and Schirk, J. and Schneider, C.M.F. and S\"odergren, L. and Tsitsilin, I. and Wallner, F. and Glaser, S.J. and Werninghaus, M. and Filipp, S.},
  journal = {Physical Review Applied},
  volume = {25},
  pages = {054037},
  year = {2026},
  doi = {10.1103/tf39-n6s3}
}

@article{Bluvstein2022,
  author = {Bluvstein, Dolev and Levine, Harry and Semeghini, Giulia and Wang, Tout T. and Ebadi, Sepehr and Kalinowski, Marcin and Keesling, Alexander and Maskara, Nishad and Pichler, Hannes and Greiner, Markus and Vuleti{\'c}, Vladan and Lukin, Mikhail D.},
  title = {{A Quantum Processor Based on Coherent Transport of Entangled Atom Arrays}},
  journal = {Nature},
  volume = {604},
  pages = {451--456},
  year = {2022},
  doi = {10.1038/s41586-022-04592-6}
}

@article{Lucero2010,
  author = {Lucero, E. and Kelly, J. and Bialczak, R. C. and Lenander, M. and Mariantoni, M. and Neeley, M. and O'Connell, A. D. and Sank, D. and Wang, H. and Weides, M. and Wenner, J. and Yamamoto, T. and Cleland, A. N. and Martinis, J. M.},
  title = {Reduced Phase Error through Optimized Control of a Superconducting Qubit},
  journal = {Physical Review A},
  volume = {82},
  pages = {042339},
  year = {2010},
  doi = {10.1103/PhysRevA.82.042339},
  url = {https://journals.aps.org/pra/abstract/10.1103/PhysRevA.82.042339},
}

@article{PRXQuantum.5.030353,
  title = {{Reducing Leakage of Single-Qubit Gates for Superconducting Quantum Processors Using Analytical Control Pulse Envelopes}},
  author = {Hyypp\"a, Eric and Veps\"al\"ainen, Antti and Papi{\v{c}}, Miha and Chan, Chun Fai and Inel, Sinan and Landra, Alessandro and Liu, Wei and Luus, J\"urgen and Marxer, Fabian and Ockeloen-Korppi, Caspar and Orbell, Sebastian and Tarasinski, Brian and Heinsoo, Johannes},
  journal = {PRX Quantum},
  volume = {5},
  pages = {030353},
  year = {2024},
  doi = {10.1103/PRXQuantum.5.030353}
}

@article{Wood2018Leakage,
  author = {Wood, Christopher J. and Gambetta, Jay M.},
  title = {Quantification and characterization of leakage errors},
  journal = {Physical Review A},
  volume = {97},
  pages = {032306},
  year = {2018},
  doi = {10.1103/PhysRevA.97.032306},
  url = {https://doi.org/10.1103/PhysRevA.97.032306},
}

@article{Magesan2012IRB,
  author = {Magesan, Easwar and Gambetta, Jay M. and Johnson, B. R. and Ryan, Colm A. and Chow, Jerry M. and Merkel, Seth T. and da Silva, Marcus P. and Keefe, George A. and Rothwell, Mary B. and Ohki, Thomas A. and Ketchen, Mark B. and Steffen, M.},
  title = {{Efficient Measurement of Quantum Gate Error by Interleaved Randomized Benchmarking}},
  journal = {Physical Review Letters},
  volume = {109},
  pages = {080505},
  year = {2012},
  doi = {10.1103/PhysRevLett.109.080505}
}

@article{Fluxonium1,
  title = {{Suppressing Counter-Rotating Errors for Fast Single-Qubit Gates with Fluxonium}},
  author = {Rower, David A. and Ding, Leon and Zhang, Helin and Hays, Max and An, Junyoung and Harrington, Patrick M. and Rosen, Ilan T. and Gertler, Jeffrey M. and Hazard, Thomas M. and Niedzielski, Bethany M. and Schwartz, Mollie E. and Gustavsson, Simon and Serniak, Kyle and Grover, Jeffrey A. and Oliver, William D.},
  journal = {PRX Quantum},
  volume = {5},
  pages = {040342},
  year = {2024},
  doi = {10.1103/PRXQuantum.5.040342}
}

@article{Fluxonium2,
  title = {{High-Coherence Fluxonium Qubit}},
  author = {Nguyen, Long B. and Lin, Yen-Hsiang and Somoroff, Aaron and Mencia, Raymond and Grabon, Nicholas and Manucharyan, Vladimir E.},
  journal = {Physical Review X},
  volume = {9},
  pages = {041041},
  year = {2019},
  doi = {10.1103/PhysRevX.9.041041}
}

@article{Bluvstein2026FaultTolerantNeutralAtomArchitecture,
  author = {Bluvstein, D. and others},
  title = {A fault-tolerant neutral-atom architecture for universal quantum computation},
  journal = {Nature},
  volume = {649},
  pages = {39--46},
  year = {2026},
  doi = {10.1038/s41586-025-09848-5}
}

@article{Park2025TransducedNoise,
  author = {Park, Jaemin and Jang, Hyeongyu and Sohn, Hanseo and Yun, Jonginn and Song, Younguk and Kang, Byungwoo and Stehouwer, Lucas E. A. and {Degli Esposti}, Davide and Scappucci, Giordano and Kim, Dohun},
  title = {Passive and active suppression of transduced noise in silicon spin qubits},
  journal = {Nature Communications},
  volume = {16},
  pages = {78},
  year = {2025},
  doi = {10.1038/s41467-024-55338-z},
  url = {https://doi.org/10.1038/s41467-024-55338-z},
}

@article{Chuang1997QPT,
  title = {Prescription for experimental determination of the dynamics of a quantum black box},
  author = {Chuang, Isaac L. and Nielsen, Michael A.},
  journal = {Journal of Modern Optics},
  volume = {44},
  number = {11-12},
  pages = {2455--2467},
  year = {1997},
  doi = {10.1080/09500349708231894},
  url = {https://doi.org/10.1080/09500349708231894},
}

@article{qutip5,
  title = {{QuTiP 5: The Quantum Toolbox in Python}},
  author = {Lambert, Neill and Gigu{\`e}re, Eric and Menczel, Paul and Li, Boxi and Hopf, Patrick and Su{\'a}rez, Gerardo and Gali, Marc and Lishman, Jake and Gadhvi, Rushiraj and Agarwal, Rochisha and Galicia, Asier and Shammah, Nathan and Nation, Paul and Johansson, J. R. and Ahmed, Shahnawaz and Cross, Simon and Pitchford, Alexander and Nori, Franco},
  journal = {Physics Reports},
  volume = {1153},
  pages = {1--62},
  year = {2026},
  doi = {10.1016/j.physrep.2025.10.001}
}

@article{sivak2026reinforcement,
  title = {Reinforcement learning control of quantum error correction},
  author = {Sivak, Volodymyr and others},
  journal = {Nature},
  pages = {879--884},
  year = {2026},
  doi = {10.1038/s41586-026-10759-2},
  volume = {655},
}

@article{sivak2023real,
  title = {Real-time quantum error correction beyond break-even},
  author = {Sivak, Volodymyr V and others},
  journal = {Nature},
  volume = {616},
  pages = {50--55},
  year = {2023},
  doi = {10.1038/s41586-023-05782-6},
}

@article{Kandala2019ErrorMitigation,
  title = {Error mitigation extends the computational reach of a noisy quantum processor},
  author = {Kandala, Abhinav and Temme, Kristan and C{\'o}rcoles, Antonio D. and Mezzacapo, Antonio and Chow, Jerry M. and Gambetta, Jay M.},
  journal = {Nature},
  volume = {567},
  pages = {491--495},
  year = {2019},
  doi = {10.1038/s41586-019-1040-7},
  url = {https://doi.org/10.1038/s41586-019-1040-7},
}

@article{Preskill2018NISQ,
  title = {Quantum Computing in the {NISQ} era and beyond},
  author = {Preskill, John},
  journal = {Quantum},
  volume = {2},
  pages = {79},
  year = {2018},
  doi = {10.22331/q-2018-08-06-79},
  url = {https://doi.org/10.22331/q-2018-08-06-79},
}

@article{Hutchings2017frequencyshift,
  title = {Tunable superconducting qubits with flux-independent coherence},
  author = {Hutchings, M. D. and Hertzberg, J. B. and Liu, Y. and Bronn, N. T. and Keefe, G. A. and Chow, J. M. and Plourde, B. L. T.},
  journal = {Physical Review Applied},
  volume = {8},
  pages = {044003},
  year = {2017},
  doi = {10.1103/PhysRevApplied.8.044003},
  url = {https://doi.org/10.1103/PhysRevApplied.8.044003},
}

@article{Rol2020Cryoscopepulse,
  title = {Time-domain characterization and correction of on-chip distortion of control pulses in a quantum processor},
  author = {Rol, M. A. and Ciorciaro, L. and Malinowski, F. K. and Tarasinski, B. M. and Sagastizabal, R. E. and Bultink, C. C. and Salath{\'e}, Y. and Haandb{\ae}k, N. and Sedivy, J. and DiCarlo, L.},
  journal = {Applied Physics Letters},
  volume = {116},
  pages = {054001},
  year = {2020},
  doi = {10.1063/1.5133894},
  url = {https://doi.org/10.1063/1.5133894},
}

@article{GoogleQuantumAI2025BelowThreshold,
  title = {Quantum error correction below the surface code threshold},
  author = {{Google Quantum AI and Collaborators}},
  journal = {Nature},
  volume = {638},
  pages = {920--926},
  year = {2025},
  doi = {10.1038/s41586-024-08449-y},
  url = {https://doi.org/10.1038/s41586-024-08449-y},
}

@article{Dennis2002TopologicalMemory,
  title = {Topological quantum memory},
  author = {Dennis, Eric and Kitaev, Alexei and Landahl, Andrew and Preskill, John},
  journal = {Journal of Mathematical Physics},
  volume = {43},
  pages = {4452--4505},
  year = {2002},
  doi = {10.1063/1.1499754},
  url = {https://doi.org/10.1063/1.1499754},
}

@article{Bravyi2024HighThresholdLDPC,
  title = {High-threshold and low-overhead fault-tolerant quantum memory},
  author = {Bravyi, Sergey and Cross, Andrew W. and Gambetta, Jay M. and Maslov, Dmitri and Rall, Patrick and Yoder, Theodore J.},
  journal = {Nature},
  volume = {627},
  pages = {778--782},
  year = {2024},
  doi = {10.1038/s41586-024-07107-7},
  url = {https://doi.org/10.1038/s41586-024-07107-7},
}

@article{Paetznick2026ImprovedLogicalErrorRates,
  title = {Improved quantum processor logical error rates via correction and detection},
  author = {Paetznick, A. and others},
  journal = {Nature},
  volume = {654},
  pages = {349--355},
  year = {2026},
  doi = {10.1038/s41586-026-10628-y}
}

@article{wang2012composite,
  title = {Composite pulses for robust universal control of singlet--triplet qubits},
  author = {Wang, Xin and Bishop, Lev S and Kestner, JP and Barnes, Edwin and Sun, Kai and Das Sarma, S},
  journal = {Nature Communications},
  volume = {3},
  pages = {997},
  year = {2012},
  doi = {10.1038/ncomms2003}
}

@article{chang2026benchmarkingdarkmattersearch,
  title = {Benchmarking dark matter search using a parity-check protocol with machine-learning optimized pulses},
  author = {Yu-Han Chang and Ilya Moskalenko and Marko Kuzmanović and Ognjen Stanisavljević and Isak Björkman and David Díez-Ibáñez and Yikun Gu and Akash V. Dixit and Igor G. Irastorza and Gheorghe Sorin Paraoanu},
  year = {2026},
  journal = {Physical Review D},
  volume = {114},
  pages = {052011},
  doi = {10.1103/m92t-xwgt}
}

@article{PhysRevA.109.012616,
  title = {{$ZZ$-interaction-free single-qubit-gate optimization in superconducting qubits}},
  author = {Watanabe, Shu and Tabuchi, Yutaka and Heya, Kentaro and Tamate, Shuhei and Nakamura, Yasunobu},
  journal = {Physical Review A},
  volume = {109},
  pages = {012616},
  year = {2024},
  doi = {10.1103/PhysRevA.109.012616}
}

@article{steph2026pra,
  title = {Optimal birobust quantum control by inverse optimization},
  author = {Hezzi, Ghaieth and Harutyunyan, Meri and Dridi, Ghassen and Mejatty, Mohamed and Gu\'erin, St\'ephane},
  journal = {Physical Review A},
  volume = {113},
  pages = {022432},
  year = {2026},
  doi = {10.1103/288v-kg8g}
}

@article{DRAG,
  title = {{Simple Pulses for Elimination of Leakage in Weakly Nonlinear Qubits}},
  author = {Motzoi, F. and Gambetta, J. M. and Rebentrost, P. and Wilhelm, F. K.},
  journal = {Physical Review Letters},
  volume = {103},
  pages = {110501},
  year = {2009},
  doi = {10.1103/PhysRevLett.103.110501}
}

@article{time-rescaling,
  title = {Time-rescaled quantum dynamics as a shortcut to adiabaticity},
  author = {de Lima Bernardo, Bert\'ulio},
  journal = {Physical Review Research},
  volume = {2},
  pages = {013133},
  year = {2020},
  doi = {10.1103/PhysRevResearch.2.013133}
}

@article{zhu2022jpb,
  title = {Robust quantum control by smooth quasi-square pulses},
  author = {Zhu, Jing-jun and Laforgue, Xavier and Chen, Xi and Gu{\'e}rin, St{\'e}phane},
  journal = {Journal of Physics B: Atomic, Molecular and Optical Physics},
  volume = {55},
  number = {19},
  pages = {194001},
  year = {2022},
  publisher = {IOP Publishing},
  doi = {10.1088/1361-6455/ac8adf},
  url = {https://doi.org/10.1088/1361-6455/ac8adf},
}

@article{composite2011prl,
  title = {{High-Fidelity Adiabatic Passage by Composite Sequences of Chirped Pulses}},
  author = {Torosov, Boyan T. and Gu\'erin, St\'ephane and Vitanov, Nikolay V.},
  journal = {Physical Review Letters},
  volume = {106},
  pages = {233001},
  year = {2011},
  doi = {10.1103/PhysRevLett.106.233001}
}

@article{NumRobust2015prl,
  title = {{Accurate and Robust Unitary Transformations of a High-Dimensional Quantum System}},
  author = {Anderson, B. E. and Sosa-Martinez, H. and Riofr\'{\i}o, C. A. and Deutsch, Ivan H. and Jessen, Poul S.},
  journal = {Physical Review Letters},
  volume = {114},
  pages = {240401},
  year = {2015},
  doi = {10.1103/PhysRevLett.114.240401},
}

@article{Auffeves1,
  title = {{Quantum Technologies Need a Quantum Energy Initiative}},
  author = {Auff{\`e}ves, A},
  journal = {PRX Quantum},
  volume = {3},
  pages = {020101},
  year = {2022},
  doi = {10.1103/PRXQuantum.3.020101}
}

@article{Auffeves2,
  title = {{Optimizing Resource Efficiencies for Scalable Full-Stack Quantum Computers}},
  author = {Fellous-Asiani, M. and Chai, J. H. and Thonnart, Y. and Ng, H. K. and Whitney, R. S. and Auff{\`e}ves, A.},
  journal = {PRX Quantum},
  volume = {4},
  pages = {040319},
  year = {2023},
  doi = {10.1103/PRXQuantum.4.040319}
}

@article{STArobust2012optimally,
  title = {Optimally robust shortcuts to population inversion in two-level quantum systems},
  author = {Ruschhaupt, Andreas and Chen, Xi and Alonso, D and Muga, JG},
  journal = {New Journal of Physics},
  volume = {14},
  pages = {093040},
  year = {2012},
  doi = {10.1088/1367-2630/14/9/093040},
}

@article{RobustSSSP,
  title = {{Robust Quantum Control by a Single-Shot Shaped Pulse}},
  author = {Daems, David and Ruschhaupt, Andreas and Sugny, Dominique and Gu{\'e}rin, St{\'e}phane},
  journal = {Physical Review Letters},
  volume = {111},
  pages = {050404},
  year = {2013},
  doi = {10.1103/PhysRevLett.111.050404}
}

@article{PRL-KAMDRAG,
  title = {{Multirobust Control Pulses Designed by Iterative Quantum Kolmogorov-Arnold-Moser Algorithm}},
  author = {Harutyunyan, Meri and Gu{\'e}rin, St{\'e}phane},
  journal = {Physical Review Letters},
  year = {2026},
  note = {Accepted for publication (23 September 2026)},
  url = {https://journals.aps.org/prl/accepted?page=2&toc_section%5B%5D=quantum-information-science-and-technology},
}

@article{Geometric2025prA,
  title = {Geometric correspondence of noisy quantum dynamics and universal robust quantum gates},
  author = {Hai, Yong-Ju and Song, Yao and Li, Junning and Zeng, Junkai and Deng, Xiu-Hao},
  journal = {Physical Review Applied},
  volume = {23},
  pages = {054002},
  year = {2025},
  doi = {10.1103/PhysRevApplied.23.054002}
}

@article{zeng2019geometric,
  title = {Geometric formalism for constructing arbitrary single-qubit dynamically corrected gates},
  author = {Zeng, Junkai and Yang, CH and Dzurak, AS and Barnes, Edwin},
  journal = {Physical Review A},
  volume = {99},
  number = {5},
  pages = {052321},
  year = {2019},
  publisher = {APS},
  doi = {10.1103/PhysRevA.99.052321},
  url = {https://doi.org/10.1103/PhysRevA.99.052321},
}

@article{Geometrical2021prxq,
  title = {Geometrical formalism for dynamically corrected gates in multiqubit systems},
  author = {Buterakos, Donovan and Das Sarma, Sankar and Barnes, Edwin},
  journal = {PRX Quantum},
  volume = {2},
  pages = {010341},
  year = {2021},
  doi = {10.1103/PRXQuantum.2.010341}
}

@article{RIOdigital2023prl,
  title = {{Digital Optimal Robust Control}},
  author = {Harutyunyan, Meri and Holweck, Fr\'ed\'eric and Sugny, Dominique and Gu\'erin, St\'ephane},
  journal = {Physical Review Letters},
  volume = {131},
  pages = {200801},
  year = {2023},
  doi = {10.1103/PhysRevLett.131.200801}
}

@article{RIO2020prl,
  title = {{Optimal Robust Quantum Control by Inverse Geometric Optimization}},
  author = {Dridi, Ghassen and Liu, Kaipeng and Gu\'erin, St\'ephane},
  journal = {Physical Review Letters},
  volume = {125},
  pages = {250403},
  year = {2020},
  doi = {10.1103/PhysRevLett.125.250403}
}

@article{Dapeng2024prl,
  title = {{Robust Quantum Gates against Correlated Noise in Integrated Quantum Chips}},
  author = {Yi, Kangyuan and Hai, Yong-Ju and Luo, Kai and Chu, Ji and Zhang, Libo and Zhou, Yuxuan and Song, Yao and Liu, Song and Yan, Tongxing and Deng, Xiu-Hao and Chen, Yuanzhen and Yu, Dapeng},
  journal = {Physical Review Letters},
  volume = {132},
  pages = {250604},
  year = {2024},
  doi = {10.1103/PhysRevLett.132.250604}
}

@article{rehammar2023low,
  title = {Low-pass filter with ultrawide stopband for quantum computing applications},
  author = {Rehammar, Robert and Gasparinetti, Simone},
  journal = {IEEE Transactions on Microwave Theory and Techniques},
  volume = {71},
  pages = {3075--3080},
  year = {2023},
  doi = {10.1109/TMTT.2023.3238543}
}

@article{biznarova2024mitigation,
  title = {Mitigation of interfacial dielectric loss in aluminum-on-silicon superconducting qubits},
  author = {Bizn{\'a}rov{\'a}, Janka and others},
  journal = {npj Quantum Information},
  volume = {10},
  pages = {78},
  year = {2024},
  doi = {10.1038/s41534-024-00868-z}
}

@article{chayanun2024characterization,
  title = {Characterization of process-related interfacial dielectric loss in aluminum-on-silicon by resonator microwave measurements, materials analysis, and imaging},
  author = {Chayanun, Lert and others},
  journal = {APL Quantum},
  volume = {1},
  year = {2024},
  doi = {10.1063/5.0208140},
  pages = {026115}
}

@misc{aggarwal2025mitigating,
  title = {Mitigating transients in flux-control signals in a superconducting quantum processor},
  author = {Aggarwal, Anuj and Fern{\'a}ndez-Pend{\'a}s, Jorge and Abad, Tahereh and Shiri, Daryoush and Jakobsson, Halld{\'o}r and Rommel, Marcus and Nylander, Andreas and Hogedal, Emil and Osman, Amr and Bizn{\'a}rov{\'a}, Janka and Rehammar, Robert and Faucci Giannelli, Michele and Fadavi Roudsari, Anita and Bylander, Jonas and Tancredi, Giovanna},
  year = {2025},
  eprint = {2503.08645},
  archivePrefix = {arXiv}
}

@article{julia-prr-2025,
  title = {Boosting projective methods for quantum process and detector tomography},
  author = {Barber\`a-Rodr\'{\i}guez, J\'ulia and Zambrano, Leonardo and Ac\'{\i}n, Antonio and Farina, Donato},
  journal = {Physical Review Research},
  volume = {7},
  pages = {013208},
  year = {2025},
  doi = {10.1103/PhysRevResearch.7.013208}
}

@article{hashim-prxquantum-2025,
  title = {{Practical Introduction to Benchmarking and Characterization of Quantum Computers}},
  author = {Hashim, Akel and Nguyen, Long B. and Goss, Noah and Marinelli, Brian and Naik, Ravi K. and Chistolini, Trevor and Hines, Jordan and Marceaux, J.P. and Kim, Yosep and Gokhale, Pranav and Tomesh, Teague and Chen, Senrui and Jiang, Liang and Ferracin, Samuele and Rudinger, Kenneth and Proctor, Timothy and Young, Kevin C. and Siddiqi, Irfan and Blume-Kohout, Robin},
  journal = {PRX Quantum},
  volume = {6},
  pages = {030202},
  numpages = {132},
  year = {2025},
  doi = {10.1103/PRXQuantum.6.030202}
}

@article{gaikwad-qst-2025,
  doi = {10.1088/2058-9565/ae0baa},
  year = {2025},
  volume = {10},
  pages = {045055},
  author = {Gaikwad, Akshay and Torres, Manuel Sebastian and Ahmed, Shahnawaz and Kockum, Anton Frisk},
  title = {Gradient-descent methods for fast quantum state tomography},
  journal = {Quantum Science and Technology}
}

@misc{Mohseni2026,
archivePrefix = {arXiv},
author = {Mohseni, Masoud and others},
eprint = {2411.10406},
title = {{How to Build a Quantum Supercomputer: Scaling from Hundreds to Millions of Qubits}},
year = {2026}
}

@article{Koch2007,
author = {Koch, J. and Yu, T. M. and Gambetta, J. and Houck, A. A. and Schuster, D. I. and Majer, J. and Blais, A. and Devoret, M. H. and Girvin, S. M. and Schoelkopf, R. J.},
doi = {10.1103/PhysRevA.76.042319},
journal = {Physical Review A},
pages = {042319},
title = {{Charge-insensitive qubit design derived from the Cooper pair box}},
volume = {76},
year = {2007}
}

@article{Abad2022,
author = {Abad, Tahereh and Fern{\'{a}}ndez-Pend{\'{a}}s, Jorge and Kockum, A. F. and Johansson, G{\"{o}}ran},
doi = {10.1103/PhysRevLett.129.150504},
journal = {Physical Review Letters},
pages = {150504},
title = {{Universal Fidelity Reduction of Quantum Operations from Weak Dissipation}},
volume = {129},
year = {2022}
}

@article{gt51-g34c,
  title = {{Qubit sensing by optimal narrowband pulses: Demonstration for Rabi frequency sensing}},
  author = {Harutyunyan, Meri and Peaudecerf, Bruno and Sugny, Dominique and Gu{\'e}rin, St{\'e}phane},
  journal = {Physical Review Research},
  volume = {7},
  pages = {043255},
  year = {2025},
  doi = {10.1103/gt51-g34c}
}

@article{4zzl-qbsf,
  title = {Neural-network-based design and implementation of fast and robust quantum gates},
  author = {Kuzmanovi{\ifmmode \acute{c}\else \'{c}\fi{}}, Marko and Moskalenko, Ilya and Chang, Yu-Han and Stanisavljevi{\ifmmode \acute{c}\else \'{c}\fi{}}, Ognjen and Warren, Christopher and Hogedal, Emil and Aggarwal, Anuj and Ahmad, Irshad and Bizn{\'a}rov{\'a}, Janka and Dahiya, Mamta and Rommel, Marcus and Nylander, Andreas and Tancredi, Giovanna and Paraoanu, Gheorghe Sorin},
  journal = {Physical Review Applied},
  volume = {26},
  pages = {034070},
  year = {2026},
  doi = {10.1103/4zzl-qbsf}
}
 
\clearpage
\onecolumngrid
\appendix

\counterwithout{equation}{section}
\counterwithout{figure}{section}
\counterwithout{table}{section}
\setcounter{page}{1}
\setcounter{section}{0}
\setcounter{subsection}{0}
\setcounter{figure}{0}
\setcounter{table}{0}
\setcounter{equation}{0}
\renewcommand{\thepage}{S\arabic{page}}
\renewcommand{\thesection}{S\arabic{section}}
\renewcommand{\thesubsection}{\thesection.\arabic{subsection}}
\renewcommand{\thesubsubsection}{\thesubsection.\arabic{subsubsection}}
\renewcommand{\thefigure}{S\arabic{figure}}
\renewcommand{\thetable}{S\arabic{table}}
\renewcommand{\theequation}{S\arabic{equation}}

\renewcommand{\theHsection}{SM.\arabic{section}}
\renewcommand{\theHfigure}{SM.\arabic{figure}}
\renewcommand{\theHtable}{SM.\arabic{table}}
\renewcommand{\theHequation}{SM.\arabic{equation}}

\begin{center}
\textbf{\large Supplemental Material for\\
``Optimal robust single-qubit gates in superconducting quantum processors''}
\end{center}

\let\SMoriginaladdcontentsline\addcontentsline
\renewcommand{\addcontentsline}[3]{%
  \ifstrequal{#1}{toc}
    {\SMoriginaladdcontentsline{smtoc}{#2}{#3}}
    {\SMoriginaladdcontentsline{#1}{#2}{#3}}%
}
\setcounter{tocdepth}{2}
\SMtableofcontents

 
\section{The robust inverse optimization pulse}
\label{sec:su2_from_tdse}

We consider a driven two-level system in the computational basis
$\{ \ket{0}, \ket{1} \}$, described by the time-dependent Hamiltonian ($\hbar\equiv 1$ here and in the following)
\begin{equation}
H_{\alpha, \delta}(t) = H_0(t) + \lambda V(t),
\label{eq:fullH}
\end{equation}
where the ideal control Hamiltonian is 
$H_{0,0}(t) = H_0(t) = [\Omega(t)\sigma_x - \Delta(t)\sigma_z]/2$, 
and $\Omega(t)$ and $\Delta(t)$ are real-valued functions of time. The perturbation
$\lambda V(t)=[\alpha \Omega(t) \sigma_x - \delta \sigma_z]/2$
describes the quasi-static control imperfections, corresponding to a relative amplitude error (pulse inhomogeneity) $\alpha$ and a detuning offset $\delta$, respectively. 

We first characterize the ideal dynamics generated by $H_0(t)$ and reconstruct the corresponding control fields through inverse trajectory engineering. We then incorporate the effect of the perturbation $\lambda V(t)$ within the perturbative framework developed in Refs.~\cite{RIO2020prl, steph2026pra} to realize single-qubit gate operations robust against simultaneous quasi-static amplitude and detuning errors while optimizing additional objectives, such as the pulse energy.

We start from the time-dependent Schr\"{o}dinger equation (TDSE) governed by the ideal Hamiltonian $H_0(t)$,
\begin{equation}
i\frac{\partial}{\partial t} \ket{\phi_0(t)} = H_0(t) \ket{\phi_0(t)} ,
\label{eq:tdse_state}
\end{equation}
whose solution,
$\ket{\phi_0(t)} = U_{0}(t) \ket{\phi_0(0)}$,
defines the evolution operator through
$i \dot{U}_{0}(t) = H_0(t) U_{0}(t)$ 
with $U_{0}(0) = \openone$.
The state can be parameterized by Bloch-sphere coordinates (the mixing angle 
$\theta(t) \equiv \theta \in [0,\pi]$, 
the relative phase 
$\varphi(t) \equiv \varphi \in [-\pi,\pi]$, 
and the global phase 
$\gamma(t) \equiv \gamma \in [0,2\pi]$)
as
\begin{equation}
\ket{\phi_0(t)} = e^{-i\frac{\gamma}{2}}
\begin{pmatrix}
e^{i\varphi/2} \cos (\theta/2) \\
e^{-i\varphi/2} \sin (\theta/2)
\end{pmatrix}.
\label{eq:state_param}
\end{equation}
Substituting \eqref{eq:state_param} into \eqref{eq:tdse_state} yields the dynamical equations
\begin{align}
\dot{\theta} &= \Omega\sin\varphi , \label{eq:theta_dot} \\
\dot{\varphi} &= \Delta+\Omega\cos\varphi\cot\theta , \label{eq:phi_dot} \\
\dot{\gamma} &= \Omega\frac{\cos\varphi}{\sin\theta} = \dot{\theta}\frac{\cot\varphi}{\sin\theta} . \label{eq:gamma_dot}
\end{align}
The first two equations govern the Bloch-sphere trajectory, while \eqref{eq:gamma_dot} determines the accumulated global phase. These equations completely characterize the ideal evolution generated by $H_0(t)$ and form the basis for reconstructing the control fields through inverse trajectory engineering.


\subsection{Quantum control via inverse trajectory engineering}

Since the ideal Hamiltonian $H_0(t)$ is determined by the control fields $\Omega(t)$ and $\Delta(t)$, the inverse-engineering procedure reconstructs these fields from a prescribed geometric trajectory in terms of $(\theta, \varphi, \gamma)$. Inverting Eqs.~(\ref{eq:theta_dot})--(\ref{eq:gamma_dot}) with respect to the control fields yields
\begin{align}
\Delta(t) &= \dot{\varphi} - \dot{\gamma}\cos\theta , \label{eq:Delta_inv}\\
\Omega(t) &= \sqrt{\dot{\theta}^2 + \dot{\gamma}^2 \sin^{2}\theta} . \label{eq:Omega_inv}
\end{align}
The relative phase $\varphi(t)$ follows geometrically from
$\dot{\theta} = \Omega\sin\varphi$ and
$\dot{\gamma} \sin\theta = \Omega \cos\varphi$, yielding
\begin{equation}
\varphi = \arctan \mleft( \frac{\dot{\theta}} {\dot{\gamma}\sin\theta} \mright) , 
\label{eq:phi_arctan}
\end{equation}
where the appropriate branch is chosen to ensure the correct quadrant. Differentiating \eqref{eq:phi_arctan} gives $\dot{\varphi}$ in terms of $\theta$, $\gamma$, and their time derivatives, which, together with \eqref{eq:Delta_inv}, completely determines the detuning $\Delta(t)$. Consequently, any smooth trajectory $(\theta,\gamma)$ satisfying the prescribed boundary conditions required to realize the target gate determines the corresponding control fields $(\Omega(t),\Delta(t))$. The control problem therefore reduces to identifying trajectories $(\theta, \gamma)$ for which the resulting control fields implement the target gate while suppressing the perturbative effects induced by $\lambda V(t)$.


\subsection{Ideal state trajectory of an $\mathbb{SU}(2)$ operator}

The evolution operator generated by the ideal Hamiltonian $H_0(t)$ can be constructed by propagating a complete orthonormal basis,
\begin{equation}
U_{0}(t)
=
\ketbra{\phi_{0}(t)}{\phi_{0}(0)}
+
|\phi^{\perp}_{0}(t)\rangle\langle\phi^{\perp}_{0}(0)|,
\end{equation}
where $\ket{\phi_0(t)}$ is the solution of the TDSE generated by $H_0(t)$ and $\ket{\phi_0^{\perp}(t)}$ is its orthogonal complement satisfying
$\braket{\phi_0^{\perp}(t)}{\phi_0(t)} = 0$.
Using the computational basis,
$\ket{\phi_0(0)} = \ket{0}$ and
$|\phi_0^{\perp}(0)\rangle = \ket{1}$,
the evolution operator can be rewritten in terms of the angles $(\theta, \varphi, \gamma)$ as
\begin{equation}
U_{0}(t) \equiv U_{0}(\theta, \varphi, \gamma) =
\begin{pmatrix}
e^{-i(\gamma-\varphi)/2}\cos(\theta/2) &
-e^{i(\gamma+\varphi)/2}\sin(\theta/2) \\
e^{-i(\gamma+\varphi)/2}\sin(\theta/2) &
e^{i(\gamma-\varphi)/2}\cos(\theta/2)
\end{pmatrix}.
\label{eq:U_param}
\end{equation}
%
To implement a target gate 
$U_{0}(t_f) = U_{\mathrm{tar}} \in \mathbb{SU}(2)$ 
at the final time $t_f$, the terminal angles $(\theta_f, \varphi_f, \gamma_f)$ must satisfy
\begin{equation}
U_{0}(\theta_f, \varphi_f, \gamma_f) = U_{\mathrm{tar}}.
\end{equation}
The initial condition is fixed by requiring
$U(0) = \openone$. 
Taking
$\ket{\phi_0(0)} = \ket{0}$
yields the initial mixing angle $\theta_i = 0$, corresponding to the north pole of the Bloch sphere. Regularity of Eqs.~(\ref{eq:theta_dot})--(\ref{eq:gamma_dot}) at $\theta_i = 0$ further requires $\cos \varphi_i = 0$, giving the initial relative phase $\varphi_i = \pi/2$, while the condition $U_{0}(0) = \openone$ gives the initial global phase $\gamma_i = \pi/2$.

As an example, the NOT family of gates is described by
\begin{equation}
U_{\mathrm{NOT};\zeta} =
\begin{pmatrix}
0 & -e^{i\zeta} \\
e^{-i\zeta} & 0
\end{pmatrix},
\qquad
\zeta = \frac{\gamma_f + \varphi_f}{2}.
\label{eq:Unot_zeta}
\end{equation}
Hence, an $X$ gate (up to an overall phase) is recovered for $\zeta=\pi/2$, corresponding to the terminal conditions
\begin{equation}
\theta_f = \pi , \qquad
\varphi_f + \gamma_f = \pi \pmod{4\pi}.
\label{eq:X_bc}
\end{equation}
Similarly, the Hadamard gate is obtained by imposing
\begin{equation}
\theta_f = \frac{\pi}{2} , \qquad
\varphi_f - \gamma_f = 0 \pmod{4\pi}.
\label{eq:H_bc}
\end{equation}
%


\subsection{Perturbative description of robustness}
\label{perturbation}
 
To quantify the robustness of the implemented gate operation, we consider the perturbed Hamiltonian in \eqref{eq:fullH}. Let $\ket{\phi_{\alpha,\delta}(t)}$ denote the solution of the TDSE generated by $H_{\alpha,\delta}(t)$, subject to the same initial condition as the ideal state, i.e.,
$\ket{\phi_{\alpha,\delta}(0)} = \ket{\phi_0(0)}$.
The deviation from the ideal evolution is characterized by the overlap amplitude
\begin{equation}
\langle \phi_0(0) | U^\dag(t,0) U_{\alpha, \delta}(t,0) | \phi_0(0) \rangle =
\braket{\phi_0(t)}{\phi_{\alpha, \delta}(t)} =
1 + O_1(t) + O_2(t) + O_3(t) + \cdots ,
\label{eq:overlap_series}
\end{equation}
where 
$O_n(t) = \mathcal{O}(|\alpha,\delta|^n)$
denotes the $n$th-order contribution in the perturbative expansion.

Rotating to the interaction picture with respect to the ideal propagator $U_0(t,0)$, the interaction Hamiltonian becomes
\begin{equation}
H^{\rm int}_{\alpha,\delta}(t) =
U_0^\dag(t,0) \lambda V(t) U_0(t,0) \equiv
\lambda V_I(t) =
\begin{pmatrix}
e(t) & f(t)\\
f^{*}(t) & -e(t)
\end{pmatrix} ,
\end{equation}
where the interaction-picture coefficients are
\begin{align}
e(t) &= \frac{1}{2} \mleft( -\delta \cos\theta + \alpha \Omega \cos\varphi \sin\theta \mright) , \\
f(t) &= \frac{1}{2} \mleft( \delta \sin\theta + \alpha \mleft[ \dot{\gamma} \sin\theta \cos\theta - i\dot{\theta} \mright] \mright) e^{i\gamma} .
\end{align}
The perturbed propagator is then expressed through the Dyson expansion
\begin{equation}
U_{\alpha,\delta}(t,0) =
U_0(t,0) \mleft[ \openone + \frac{\lambda}{i}W_1(t) + \mleft( \frac{\lambda}{i} \mright)^2 W_2(t) + \cdots \mright] ,
\label{eq:dyson_U}
\end{equation}
where
\begin{equation}
\lambda^n W_n(t)=
\int_{0}^{t} dt_0 V_I(t_0)
\int_{0}^{t_0} dt_1 V_I(t_1)
\cdots
\int_{0}^{t_{n-2}} dt_{n-1} V_I(t_{n-1}).
\label{eq:Wn_def}
\end{equation}
Projecting the Dyson terms onto the computational basis $\{ \ket{0}, \ket{1} \}$, we define
\begin{equation}
O_n(t) = \mleft( \frac{\lambda}{i} \mright)^n \langle 0 | W_n(t) | 0 \rangle , \qquad
Q_n(t) = \mleft( \frac{\lambda}{i} \mright)^n \langle 1 | W_n(t) | 0 \rangle ,
\label{eq:On_Qn}
\end{equation}
where $O_n$ and $Q_n$ characterize the $n$th-order phase correction along the target branch and the transition amplitude to the orthogonal state, respectively.

The performance of the implemented gate can be quantified by the trace fidelity,
\begin{equation}
F_{\rm gate} =
\frac{1}{2} \mleft| \mathrm{Tr} \mleft( U_0^\dag(t_f) U_{\alpha, \delta}(t_f) \mright) \mright| =
\Re[\braket{\phi_0(t_f)}{\phi_{\alpha, \delta}(t_f)}] .
\label{eq:trace_fidelity}
\end{equation}
Robust gate operation is therefore achieved by designing the geometric trajectory $(\theta, \gamma)$ (and hence the corresponding control fields) such that the leading perturbative corrections vanish at the final time while preserving the target evolution
$U_{\alpha,\delta}(t_f)=U_{\rm tar}$ 
up to a global phase. In particular, first- and second-order robustness require
$\Re[{O}_1(t_f)] = 0$ and
$\Re[{O}_1(t_f)] = \Re[{O}_2(t_f)] = 0$,
respectively.

Up to second order, these conditions reduce to the integral constraints
\begin{equation}
\int_{0}^{t_f}e(t) dt = 0 , \qquad
\int_{0}^{t_f}(t) dt = 0 ,
\label{eq:gate_robust_core}
\end{equation}
which are equivalent to six independent real constraints (denoted by $\psi_j = 0$) on the geometric trajectory, as derived in Ref.~\cite{steph2026pra}:
\begin{align}
\psi_1(\theta) &= \int_{t_i}^{t_f} \cos\theta(t) dt = 0 , \label{eq:psi1} \\
\psi_2(\dot{\gamma}, \theta) &= \int_{t_i}^{t_f} \dot{\gamma}(t) \sin^2\theta(t) dt = \int_{\gamma_i}^{\gamma_f} \sin^2\theta d\gamma=0, \label{eq:psi2} \\
\psi_3(\gamma, \theta) &= \int_{t_i}^{t_f} \sin\theta(t) \sin\gamma(t) dt = 0, \label{eq:psi3} \\
\psi_4(\gamma, \theta) &= \int_{t_i}^{t_f} \sin\theta(t) \cos\gamma(t) dt = 0, \label{eq:psi4} \\
\psi_5(\gamma, \theta, \dot\theta) &= \int_{t_i}^{t_f} \dot\theta(t) \cos\gamma(t) \sin^2\theta(t) dt = \int_{\theta_i}^{\theta_f} \cos\gamma \sin^2\theta d\theta = 0 , \label{eq:psi5} \\
\psi_6(\gamma, \theta, \dot\theta) &= \int_{t_i}^{t_f} \dot\theta(t) \sin\gamma(t) \sin^2\theta(t) dt = \int_{\theta_i}^{\theta_f} \sin\gamma \sin^2\theta d\theta = 0 . \label{eq:psi6}
\end{align}
For $\psi_5$ and $\psi_6$, integration by parts gives explicit boundary expressions, such that these constraints encode both bulk and endpoint information.


\subsection{Energy-optimal robust control}

In addition to satisfying the gate and robustness conditions, we seek control fields that minimize the pulse energy. Using \eqref{eq:Omega_inv}, the corresponding cost functional is
\begin{equation}
\mathcal{E} =
\int_{0}^{t_f} \Omega^2(t) dt =
\int_{0}^{t_f} \mleft( \dot{\theta}^2 + \dot{\gamma}^2 \sin^2\theta \mright) dt =
\int_{0}^{t_f} \mathcal{L}_{0} dt ,
\label{eq:energy_cost}
\end{equation}
where $\mathcal{L}_{0}$ is the unconstrained Lagrangian associated with the pulse energy.
To determine the energy-optimal robust trajectory $(\theta, \gamma)$, we augment the Lagrangian by incorporating the robustness constraints through Lagrange multipliers:
\begin{equation}
\mathcal{L}(\theta, \gamma, \dot{\theta}, \dot{\gamma}) =
\mathcal{L}_{0} + \sum_{j=1}^{6} \lambda_j \varphi_j ,
\label{eq:aug_L}
\end{equation}
where $\lambda_j$ are constant Lagrange multipliers associated with the six independent robustness constraints, written as
$\psi_j = \int_{0}^{t_f} \varphi_j(t) dt = 0$,
where $j$ takes an integer value in $[1,6]$. The optimal trajectory is obtained from the constrained Euler--Lagrange equations
\begin{equation}
\frac{\partial\mathcal{L}}{\partial\eta} -
\frac{d}{dt} \mleft( \frac{\partial\mathcal{L}}{\partial\dot{\eta}} \mright)
= 0 , \qquad
\eta \in \{\theta,\gamma\},
\end{equation}
which yield the coupled second-order differential equations
\begin{align}
&\ddot{\gamma} \sin\theta
+ 2 \dot{\gamma} \dot{\theta} \cos\theta
- (\lambda_4 \cos\gamma - \lambda_3 \sin\gamma)
+ \dot{\theta} \sin\theta (\lambda_1 \sin\gamma - \lambda_2 \cos\gamma)
+ \lambda_6 \dot{\theta} \cos\theta
= 0 , \label{eq:EL_gate_gamma} \\
&\ddot{\theta}
- \dot{\gamma}^2 \sin\theta \cos\theta
- \cos\theta (\lambda_3 \cos\gamma + \lambda_4 \sin\gamma)
- \dot{\gamma} \sin^2\theta (\lambda_1 \sin\gamma - \lambda_2 \cos\gamma)
+ \lambda_5 \sin\theta
- \lambda_6 \dot{\gamma} \sin\theta \cos\theta
= 0 . \label{eq:EL_gate_theta}
\end{align}
%

\begin{figure} 
\centering
\includegraphics[width=1\columnwidth]{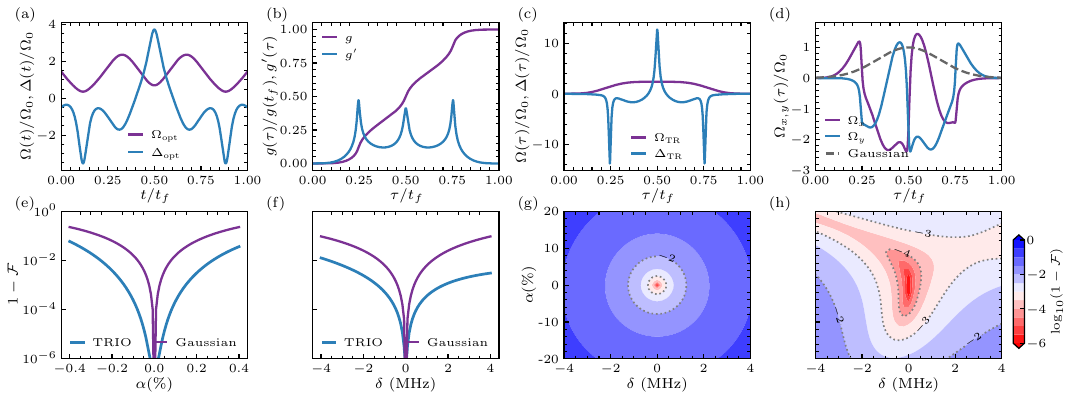}
\caption{\textbf{Simulated TRIO pulse for an X gate in a two-level system.}
\textbf{a}, The dimensionless energy-optimal pulses ${\Omega_{\rm opt}, \Delta_{\rm opt}}$ for a two-level system. 
\textbf{b}, The time-rescaling function rescales the dimensionless optimal pulse into 
\textbf{c}, a hardware-efficient pulse envelope. 
\textbf{d}, The phase-frame pulses for a frequency-fixed qubit compared with a Gaussian envelope (gray dashed curve).
\textbf{e}, \textbf{f}, Robustness simulations: infidelity as a function of amplitude deviation $\alpha$ and detuning $\delta$, respectively, for TRIO (blue) and Gaussian (purple) pulses.
\textbf{g}, \textbf{h}, Infidelity as a function of the combined error for the Gaussian and the TRIO pulse, respectively. 
Note that the unit $\Omega_{0}  = \max[{\Omega_{\rm G}(t)}]$ refers to the maximum value of a Gaussian $\pi$-pulse, e.g., the dashed curve in \textbf{d}.
\label{fig:robust-two-level}}
\end{figure}

Solving Eqs.~(\ref{eq:EL_gate_gamma}) and (\ref{eq:EL_gate_theta}) together with the gate boundary conditions and the six robustness constraints determines the Lagrange multipliers $\lambda_j$ and, consequently, the energy-optimal geometric trajectory $(\theta, \gamma)$. The corresponding optimal Rabi frequency $\Omega^{\rm opt}(t)$ is reconstructed from \eqref{eq:Omega_inv}, whereas the detuning is obtained as
\begin{align}
\Delta^{\rm opt}(t) &= \sin\theta(\lambda_1\sin\gamma-\lambda_2\cos\gamma) \nonumber \\
&+ \frac{ \dot{\theta} (\lambda_3 \sin\gamma - \lambda_4 \cos\gamma) 
+ \dot{\gamma} \cos\theta \sin\theta (\lambda_3 \cos\gamma + \lambda_4 \sin\gamma)
- \lambda_5 \dot{\gamma} \sin^2\theta}{\Omega^2}
- \lambda_6 \cos\theta .
\label{eq:Delta_gate}
\end{align}
The optimized control fields $(\Omega^{\rm opt}(t), \Delta^{\rm opt}(t))$ define the bi-robust inverse optimization (bi-RIO) pulse introduced in Ref.~\cite{steph2026pra}. As an example, the energy-optimal bi-RIO solution for a $X$ gate has a minimum pulse energy of $\mathcal{E}^{\rm opt}_{\rm X} \simeq 3.461\pi$, attained for
$\{ \lambda_j \} = \{ 1.19250104\Omega_{0} , -0.8884966\Omega_{0}, -0.34026348\Omega^{2}_{0} , -0.82990986\Omega^{2}_{0} , 0, 0 \}$,
while satisfying all six robustness constraints ($\psi_i \le 10^{-8}$) and the prescribed gate boundary conditions, as shown in \figpanel{fig:robust-two-level}{a}. 



\section{The TRIO pulse}
\label{sec:crab_opt}

Since the optimized bi-RIO pulse generally exhibits nonvanishing derivatives at its boundaries, it is not directly suitable for the derivative removal by adiabatic gate (DRAG)~\cite{DRAG} correction required to suppress leakage in the weakly anharmonic three-level system considered later. To overcome this limitation, we introduce a time-rescaling (TR) procedure that transforms the optimized bi-RIO controls $\{\Omega^{\rm opt}(t), \Delta^{\rm opt}(t)\}$ into a smooth control protocol suitable for subsequent DRAG correction. Specifically, we introduce a monotonic time-rescaling function $g(\tau)$ satisfying
\begin{equation}
t=g(\tau) , \qquad
g(0)=0 , \quad
g(\tau_f)=t_f,
\end{equation}
with derivative
\begin{equation}
g'(\tau)=\frac{dt}{d\tau}.
\end{equation}
Under this transformation, the time-dependent Schr\"{o}dinger equation governed by the perturbed bi-RIO Hamiltonian $H_{\alpha, \delta}^{\rm opt}$ becomes
\begin{equation}
i\frac{d}{d\tau}\ket{\phi_{\alpha,\delta}(\tau)} =
g'(\tau) H_{\alpha, \delta}^{\rm opt} (g(\tau)) \ket{\phi_{\alpha, \delta}(\tau)} \equiv
H_{\alpha, \delta}^{\rm TR}(\tau) \ket{\phi_{\alpha, \delta}(\tau)} ,
\end{equation}
where the effective Hamiltonian in the rescaled time frame is
\begin{equation}
\begin{aligned}
H_{\alpha,\delta}^{\rm TR}(\tau) &= \frac{g'(\tau)}{2} 
\mleft[ - \mleft( \Delta^{\rm opt}(g(\tau)) + \delta \mright) \sigma_z
+ (1 + \alpha) \Omega^{\rm opt} (g(\tau)) \sigma_x \mright] \\
&= \frac{1}{2}
\mleft[ - \mleft( \Delta_{\rm TR}(\tau) + g'(\tau) \delta \mright) \sigma_z
+ (1 + \alpha) \Omega_{\rm TR}(\tau) \sigma_x \mright] .
\end{aligned}
\label{Htau}
\end{equation}
The rescaled control fields are therefore
\begin{equation}
\Omega_{\rm TR}(\tau) = g'(\tau) \Omega^{\rm opt}(g(\tau)) , \qquad
\Delta_{\rm TR}(\tau) = g'(\tau) \Delta^{\rm opt}(g(\tau)) .
\end{equation}

In the absence of perturbations, the TR pulse implements the same target $X$ gate as the original energy-optimal bi-RIO pulse while matching its Rabi-energy cost. In practice, the rescaled Rabi frequency $\Omega_{\rm TR}(\tau)$ is initialized using a hyper-Gaussian {pulse}~\cite{time-rescaling, zhu2022jpb},
\beq
\Omega_{\rm TR}(\tau)=A\exp\mleft[-\mleft( \frac{\tau-t_f/2}{2\sigma}\mright)^{2n}\mright] ,
\label{eq:hyper_gauss_appendix}
\eeq
where we choose $n=2$ through the whole numerical simulation and experimental implementation to prevent high-power-induced leakage. Therefore, the corresponding time-rescaling function $g(\tau)$ is determined, as shown in \figpanel{fig:robust-two-level}{b}. The rescaled detuning $\Delta_{\rm TR}(\tau)$ is then reconstructed accordingly, yielding the smooth control fields shown in \figpanel{fig:robust-two-level}{c}. 

Having established the time-rescaling protocol and the corresponding reconstructed controls, we now return to the original time variable $t$ to describe their experimental implementation. For a fixed-frequency transmon, the reconstructed time-dependent detuning is implemented through microwave phase modulation rather than by directly tuning the qubit frequency. The laboratory-frame Hamiltonian of the two-level system, including quasi-static control errors, is
\begin{equation}
H_{\rm lab}(t) =
\frac{1}{2} \mleft[ (\omega_q - \delta) \sigma_z
+ 2 (1 + \alpha) \Omega(t) \cos(\omega_d t + \phi_{\rm eff}(t)) \sigma_x \mright] ,
\end{equation}
where $\omega_q$ and $\omega_d$ denote the qubit and drive frequencies, respectively, while $\alpha$ and $\delta$ represent quasi-static amplitude and frequency errors. Transforming to a rotating frame via
$U(t)=\exp(-i \omega_q t \sigma_z / 2)$
and applying the rotating-wave approximation yields
\begin{equation}
H_{\rm lab}^{R}(t) =
\frac{1}{2} \left[-\delta \sigma_z +
(1+\alpha)\Omega(t)e^{-i[(\omega_d-\omega_q)t+\phi_{\rm eff}(t)]}\sigma_{+}
+\mathrm{H.c.}\right] .
\end{equation}
The corresponding state satisfies the TDSE
\begin{equation}
i\frac{d}{dt}\ket{\psi_{\alpha, \delta}(t)}=H_{\rm lab}^{R}(t)\ket{\psi_{\alpha, \delta}(t)},
\end{equation}
with the evolution operator defined by 
$\ket{\psi_{\alpha, \delta}(t)}=U_{\rm lab}^{R}(t) \ket{\psi_{\alpha, \delta}(0)}$.

Under the resonant driving condition $\omega_d = \omega_q$, the reconstructed time-dependent detuning is encoded entirely in the accumulated phase modulation of the microwave drive. Accordingly, the reconstructed controls are implemented through
\begin{equation}
\Omega(t)=\Omega^{\rm opt}(t) , \qquad
\Delta^{\rm opt}(t)=\dot{\phi}_{\rm eff}(t),\qquad
\phi_{\rm eff}(t)=\int_0^t\Delta^{\rm opt}(s) ds .
\end{equation}
To recover the effective Hamiltonian in \eqref{eq:fullH}, we introduce an additional time-dependent unitary transformation,
\begin{equation}
\ket{\psi_{\alpha, \delta}(t)} = W(t) \ket{\phi_{\alpha, \delta}(t)} ,
\label{eq:rotation}
\end{equation}
where
$W(t) = e^{-i \phi_{\rm eff}(t) \sigma_z / 2}$.
The resulting transformed Hamiltonian is
\begin{align}
H_{\alpha,\delta}(t) 
&= W^\dag(t) H_{\rm lab}^{R}(t) W(t) - i W^\dag(t) \dot{W(t)} \nonumber \\
&= \frac{1}{2} \mleft[ - \mleft( \Delta^{\rm opt}(t) + \delta \mright) \sigma_z
+ (1 + \alpha) \Omega^{\rm opt}(t) \sigma_x \mright] ,
\end{align}
which reproduces the effective Hamiltonian in \eqref{eq:fullH}. The evolution operators in the two frames are related by
\begin{equation}
U_{\rm lab}^{R}(t) = W(t) U_{\alpha, \delta}(t) W^\dag(0).
\end{equation}
Since $\phi_{\rm eff}(0) = 0$, we have $W(0) = I$. Thus, in the absence of control errors, the target gate $U_0(t_f)$ generated by the reconstructed controls is related to the experimentally implemented rotating-frame propagator through
\begin{equation}
U_0(T) = W^\dag(T) U_{\rm lab}^{R}(T) \big|_{\alpha = \delta = 0} .
\end{equation}

Applying the laboratory-frame implementation described above to the smooth time-rescaled controls, we obtain the experimentally realizable TRIO pulse, expressed as the complex microwave envelope
\beq
\Omega_{\rm TRIO}(\tau) =
e^{i\int_{0}^{\tau} \Delta_{\rm TR}(s) ds} \Omega_{\rm TR}(\tau) =
\Omega_x(\tau) + i \Omega_y(\tau) ,
\eeq
where
$\Omega_x(\tau) = \Re[\Omega_{\rm TRIO}(\tau)]$
and
$\Omega_y(\tau) = \Im[\Omega_{\rm TRIO}(\tau)]$
are the two microwave quadratures, as shown in \figpanel{fig:robust-two-level}{d}. The resulting pulse preserves the target gate propagator up to the additional phase rotation associated with the transformation $W(t_f)$.

Although the TR transformation preserves the target gate operation, it also rescales the perturbation, such that the effective detuning error becomes
$g'(\tau)\delta$.
Consequently, the robustness of the original bi-RIO pulse is preserved only in the rescaled perturbative frame rather than under the physical error model, where the detuning shift remains constant throughout the gate operation. As a result, the robustness against detuning errors is partially degraded after time rescaling. Nevertheless, the reconstructed pulse remains substantially more robust than a conventional $\pi$ pulse while satisfying the smooth-boundary conditions required for DRAG implementation, as shown in \figpanels{fig:robust-two-level}{e}{h}. Moreover, the pulse energy can be preserved under the time-rescaling (TR) transformation by appropriately choosing the rescaling function $g(\tau)\approx\tau$ such that  
{\textcolor{black}{$\int_0^{t_f}\Omega_{\rm TR}^2(\tau)d\tau
=\int_0^{t_f} g'(\tau)\left[\Omega^{\rm opt}(g(\tau))\right]^2d g(\tau)\approx\int_0^{t_f}[\Omega^{\rm opt}(g(\tau))]^{2}dg(\tau)$. }}

Furthermore, the unitary transformation $W(t)$ introduced in \eqref{eq:rotation} gives rise to an additional phase rotation that must be compensated to recover the target gate propagator in the absence of control errors. This compensation is implemented experimentally through a virtual-$Z$ (Vz) gate during the calibration procedure, as explained in the following sections.


\section{Leakage mitigation: DRAG quadrature and Stark-shift compensation}

Having constructed the experimentally realizable TRIO pulse from the smooth time-rescaled controls, we now extend the ideal two-level description to more levels to better describe superconducting transmon qubits. The finite anharmonicity in these qubits gives rise to unwanted population leakage from the computational subspace during fast gate operations. To mitigate this effect, we apply the DRAG technique~\cite{DRAG} to the smooth time-rescaled controls underlying the TRIO pulse. This technique suppresses nonadiabatic excitation of higher energy levels through an additional derivative quadrature together with a compensation of the AC Stark shift.

To account for leakage, we consider the lowest three energy levels of the transmon. In the basis $\{ \ket{0}, \ket{1}, \ket{2} \}$ and the rotating frame of the microwave drive, the three-level Hamiltonian is
\begin{equation}
\frac{H_3(\tau)}{\hbar} =
\begin{pmatrix}
0 & s^{*}(\tau)/2 & 0 \\
s(\tau)/2 & \Delta(\tau) & \sqrt{2} s^{*}(\tau)/2 \\
0 & \sqrt{2} s(\tau)/2 & \xi + 2 \Delta(\tau)
\end{pmatrix},
\label{eq:D_three_level_drag_frame}
\end{equation}
where
$s(\tau) = \Omega^{s}_x(\tau) + i\Omega^{s}_y(\tau)$
is the complex microwave envelope, with $\Omega^{s}_x(\tau)$ and $\Omega^{s}_y(\tau)$ denoting the in-phase and quadrature components of the drive, respectively. Here, $\xi = \omega_{12} - \omega_{01} < 0$ is the transmon anharmonicity. The factor $\sqrt{2}$ arises from the harmonic oscillator-like matrix elements of the transmon ladder operator. Before applying DRAG, we work in the effective frame in which the phase modulation associated with the TRIO pulse has been removed. In this frame, the uncorrected time-rescaled controls are
\begin{equation}
\Omega^{s}_x(\tau) = \Omega_{\rm TR}(\tau) , \qquad
\Omega^{s}_y(\tau) = 0 , \qquad
\Delta(\tau) = \Delta_{\rm TR}(\tau) .
\end{equation}
%

\begin{figure}
\centering
\includegraphics[width=1\columnwidth]{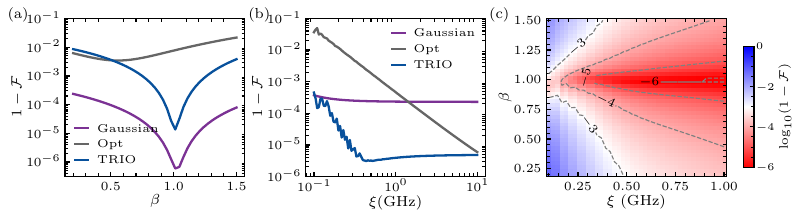}
\caption{\textbf{Numerical simulation of leakage suppression.}
\textbf{a}, Gate infidelity as a function of the DRAG strength $\beta$ defined in \eqref{eq:D_drag_complex_envelope}, for the Gaussian and TRIO pulses, along with a comparison to the pulse optimized for a two-level system. 
\textbf{b}, Gate infidelity as a function of anharmonicity $\xi$ in the presence of errors $\delta = \SI{0.1}{\mega\hertz}$ and $\alpha = \SI{1}{\percent}$. \textbf{c}, Gate infidelity as a function of both $\beta$ and $\xi$ for the TRIO pulse. Other parameters: $T_{\rm gate}= \SI{90}{\nano\second}$. The dynamics are described by the three-level Hamiltonian defined in \eqref{eq:D_three_level_drag_frame}.
\label{fig:robust-three-level}}
\end{figure}

\begin{figure}
\centering
\includegraphics[width=0.9\columnwidth]{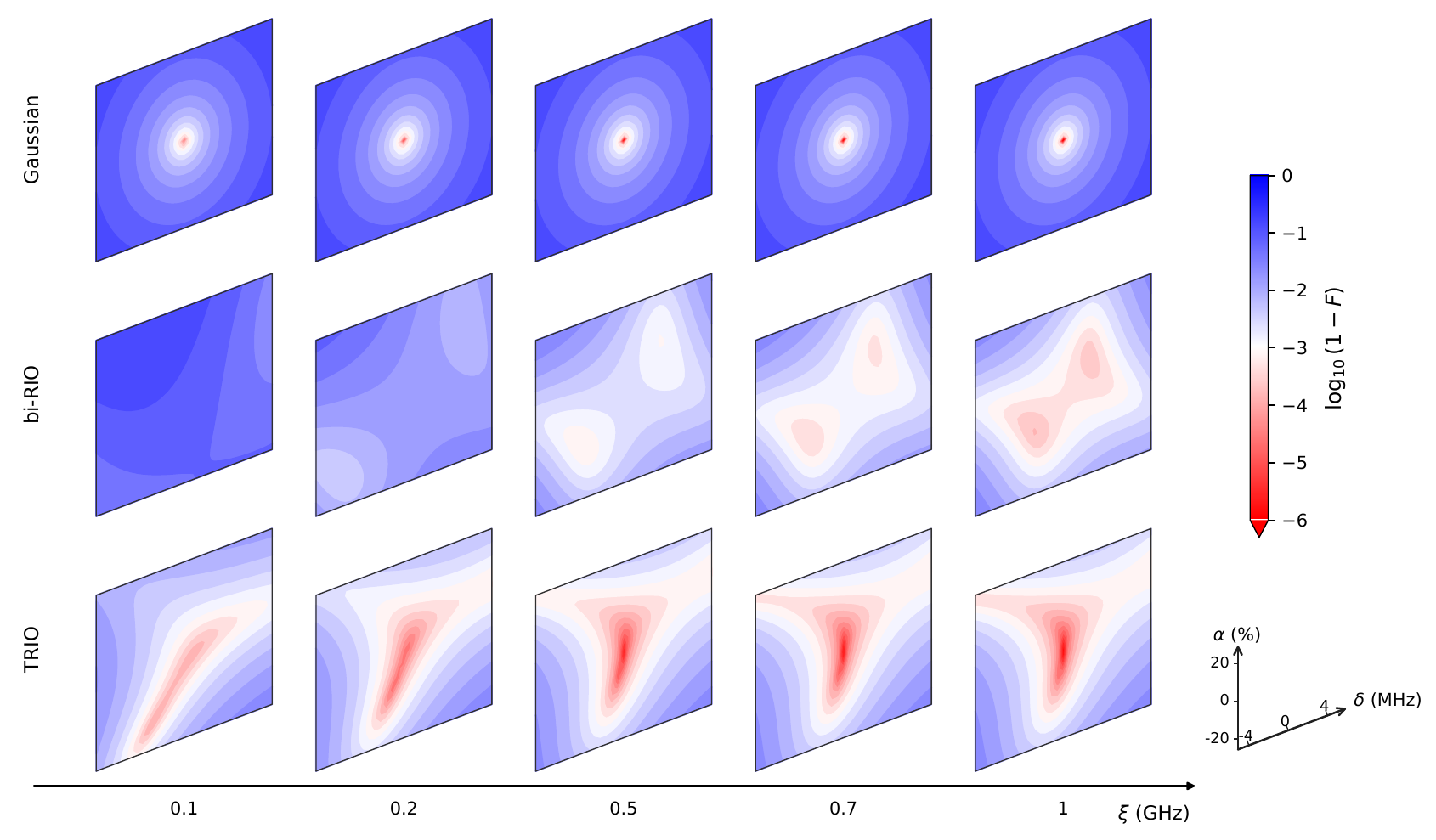}
\caption{\textbf{Numerical simulation of robustness with leakage errors.}
Using the same error setting $\delta \in [-4,4]\,\text{MHz}$ and $\alpha \in [-20,20]\,\%$ as in the main text, the plots show the robustness behavior as a function of the anharmonicity $\xi$ for the Gaussian, bi-RIO, and TRIO pulses, respectively. All three pulses are corrected by DRAG with the parameter $\beta = 1$. The control pulse called ``bi-RIO'' is the optimal result for a two-level system, proposed in Ref.~\cite{steph2026pra}.
\label{fig:robust-three-level-scan}}
\end{figure}

Treating the drive strength as a small parameter,
$\epsilon = \mleft| \max \{\Omega_{\rm TR}\} / \xi \mright| \ll 1$,
we derive an effective two-level Hamiltonian by performing a perturbative Schrieffer--Wolff (SW) transformation,
\begin{equation}
H_{\rm eff} =
e^{-\epsilon S} H_3 e^{\epsilon S}
- i e^{-\epsilon S} \partial_t e^{\epsilon S} ,
\end{equation}

The anti-Hermitian generator $S(t)$ is expanded perturbatively in powers of $\epsilon$ as $S(t) = S_1(t) + \epsilon S_2(t) + \cdots$,
where $S_n(t) = \sum_{\substack{i,j\in\{0,1,2\}\\i<j}}
    \left[
    s_{ij}^{n}(t)|i\rangle\langle j|
    - s_{ij}^{n*}(t)|j\rangle\langle i|
    \right]$.
Here, $s_{ij}^{n}(t)$ are tunable, order-dependent coefficients that can be optimized to suppress leakage at $n-$th order of the perturbative expansion. The generator is constructed to eliminate the coupling between the computational and leakage subspaces to first order in $\epsilon$, thereby recovering the population dynamics of an effective two-level system up to $\mathcal{O}(\epsilon)$. 

Expanding the transformed Hamiltonian to first order in $\epsilon$ yields the DRAG conditions
\begin{equation}
\Omega_y^{\rm DRAG}(\tau) = - \frac{\dot{\Omega}_{\rm TR}(\tau)}{\xi} , \qquad
\Omega_x^{\rm DRAG}(\tau) = \Omega_{\rm TR}(\tau) - \frac{\Omega_{\rm TR}(\tau) \Delta_{\rm TR}(\tau)}{\xi},
\end{equation}
which combine to give the complex DRAG envelope
\begin{equation}
s_{\rm DRAG}(\tau) =
\Omega^{\rm opt}(\tau) -
\frac{\Omega_{\rm TR}(\tau) \Delta_{\rm TR}(\tau)}{\xi} -
i \beta \frac{\dot{\Omega}_{\rm TR}(\tau)}{\xi} ,
\label{eq:D_drag_complex_envelope}
\end{equation}
where $\beta$ is an experimentally calibrated scaling factor for the derivative quadrature. The optimal value of $\beta$ is determined by maximizing the gate fidelity in the absence of control errors, as depicted in \figref{fig:robust-three-level}. The resulting DRAG pulse suppresses the leading-order leakage coupling to the state $\ket{2}$. The numerical results show that TRIO gate errors can reach below $10^{-6}$ for an anharmonicity of \SI{1}{\giga\hertz}; see \figpanel{fig:robust-three-level}{b}.

Besides leakage, the off-resonant coupling to $\ket{2}$ also induces an AC Stark shift of the computational transition. Eliminating this shift perturbatively to first order in $\epsilon$ gives
\begin{equation}
\delta_{\rm Stark}(\tau) = \frac{\mleft| \Omega_{\rm TR}(\tau) \mright|^2}{2 \xi} ,
\label{eq:D_stark_shift_from_leakage}
\end{equation}
which shifts the effective qubit transition frequency. This shift is compensated by modifying the target detuning according to
\begin{equation}
\Delta_{\rm DRAG}(\tau) = \Delta_{\rm TR}(\tau) - \delta_{\rm Stark}(\tau) ,
\label{eq:D_leakage_mitigated_detuning}
\end{equation}
yielding the DRAG-corrected control fields.

Finally, following the laboratory-frame phase-encoding procedure introduced above, we express the DRAG-corrected time-rescaled controls as
$\tilde{\Omega}_{\rm TR}(\tau) \equiv s_{\rm DRAG}(\tau)$
and
$\tilde{\Delta}_{\rm TR}(\tau) \equiv \Delta_{\rm DRAG}(\tau)$,
consistent with the notation used in the main text. The resulting DRAG-corrected TRIO pulse is then implemented through the complex microwave envelope
\begin{equation}
\tilde{\Omega}_{\rm TRIO}(\tau) = e^{i\tilde{\phi}_{\rm eff}(\tau)} \tilde{\Omega}_{\rm TR}(\tau) ,
\end{equation}
where the accumulated phase is given by
\begin{equation}
\tilde{\phi}_{\rm eff}(\tau) = \int_0^\tau \tilde{\Delta}_{\rm TR}(s) ds ,
\end{equation}
which corresponds to \eqref{eq:TRIO} in the main text. This construction incorporates the DRAG corrections into the time-rescaled protocol, yielding an experimentally realizable complex microwave envelope that provides leakage suppression and AC Stark-shift compensation. The robustness against leakage errors is evaluated and compared for the DRAG-corrected TRIO and Gaussian $\pi$-pulse derived within the three-level framework, as well as the bi-RIO protocol obtained within the two-level framework, as shown in Fig. \ref{fig:robust-three-level-scan}.


\begin{figure*} 
\centering
\includegraphics[width=0.6\columnwidth]{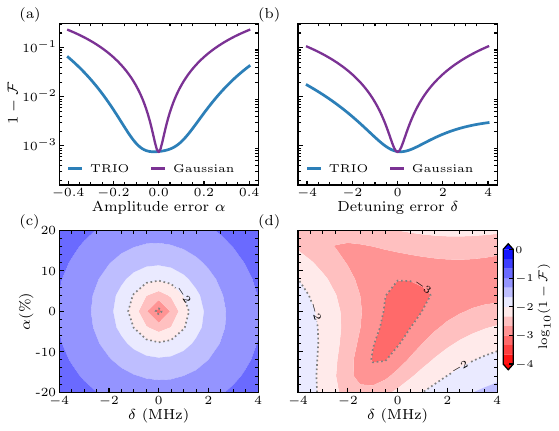}
\caption{\textbf{Robustness simulation with realistic parameters.}
\textbf{a}, \textbf{b}, Gate infidelity as functions of amplitude and detuning error, respectively, for the Gaussian and TRIO pulses. 
\textbf{c}, \textbf{d}, Gate infidelity as a function of both amplitude and detuning error for the Gaussian and TRIO pulses, respectively.
The gate fidelity is simulated with a five-level system using the master equation by taking into account finite coherence with $T_1 = \SI{75}{\micro\second}$ and $T_2^* = \SI{55}{\micro\second}$. Other parameters: DRAG coefficient $\beta = 1$, anharmonicity $\xi = \SI{239}{\mega\hertz}$\textcolor{red}{}. 
\label{fig:robust-three-level_derio_open}}
\end{figure*}

We simulate the driven single-qubit gate dynamics using a truncated transmon model. For the multilevel simulation, we use a Kerr oscillator truncated to $d$ levels,
\begin{equation}
H_d(t) = \frac{\xi}{2} a^\dag a^\dag a a + \frac{1}{2} s(t) a^\dag + \frac{1}{2} s^*(t) a + \Delta(t) a^\dag a .
\end{equation}
Here, $\xi$ is the anharmonicity and $a$ ($a^\dag$) is the annihilation (creation) operator in the truncated Hilbert space. Amplitude and frequency errors are included as
\begin{equation}
s'(t) \rightarrow (1 + \alpha)s(t) , \qquad
\Delta'(t) \rightarrow \Delta(t) + \delta .
\end{equation}
The time-dependent coefficients are interpolated using cubic splines, and the dynamics are propagated using QuTiP~\cite{qutip5}.
To include decoherence, we solve the Lindblad master equation
\begin{equation}
\dot{\rho} = -i [H(t), \rho] + \sum_j \mathcal{D}[c_j] \rho ,
\end{equation}
with jump operator $\mathcal{D}[c]\rho = c \rho c^\dag - \frac{1}{2} \{c^\dag c, \rho\}$.
Energy relaxation and pure dephasing are included through $\Gamma_1 = 1 / T_1$ and $c_\phi = \sqrt{2 \Gamma_\phi} a^\dag a$, where $\Gamma_\phi = \frac{1}{T_2^*} - \frac{1}{2T_1}$.
Thus, the $T_2^*$ measured in Ramsey experiments is treated as the effective total dephasing time. The gate performance of the Gaussian and TRIO pulses in the presence of finite coherence times is compared in \figref{fig:robust-three-level_derio_open}.

For closed-system simulations, we compute the propagator $U(t)$ directly and extract the computational block $M$ from the final propagator. A virtual-$Z$ phase compensation can be applied as $W(\phi) = \exp(-i \phi a^\dag a)$.
For an ideal $X$ gate, the average gate fidelity is evaluated as~\cite{nielsen2002simple}
\begin{equation}
\mathcal{F}_{\mathrm{avg}} = \frac{\mathrm{Tr}(M^\dag M) + \mleft| \mathrm{Tr}(X^\dag M) \mright|^2}{6} .
\end{equation}
For open-system simulations, the gate is treated as a quantum channel. We propagate the six input states $\ket{0}$, $\ket{1}$, $\ket{+}$, $\ket{-}$, $\ket{+i}$, and $\ket{-i}$, compare each final density matrix with the ideal $X$-gate output, and average the six state fidelities. Leakage is computed from the simulated populations as
\begin{equation}
L(t) = 1 - P_0(t) - P_1(t) ,
\end{equation}
which reduces to $L(t) = P_2(t)$ for a three-level truncation.

 
\section{Universal gate sets with TRIO pulses}

In order to implement the conventional Hadamard gate, we adopt the SU(2) Hadamard-type construction of Ref.~\cite{steph2026pra}. The gate is parameterized as
\begin{equation}
U_{\mathrm{H};\zeta,\eta} = \frac{1}{\sqrt{2}}
\begin{pmatrix}
e^{i\eta} & -e^{i\zeta} \\
e^{-i\zeta} & e^{-i\eta}
\end{pmatrix}, \qquad
\eta = \frac{\varphi_f - \gamma_f}{2} , \qquad
\zeta = \frac{\varphi_f + \gamma_f}{2} .
\end{equation}
Following Ref.~\cite{steph2026pra}, we choose the boundary conditions
$\theta_i = 0$, $\theta_f = \pi/2$, and $\gamma_i = \varphi_i = \pi/2$.
The corresponding numerical solution for Lagrange multipliers (c.f. \eqref{eq:aug_L}) are
$\lambda_1 = 0.0918\Omega_{0} $, $\lambda_2 = -2.1038\Omega_{0} $,
$\lambda_3 = 0.5300\Omega^{2}_{0}  $, $\lambda_4 = -0.5994\Omega^{2}_{0}  $,
$\lambda_5 = -0.8001\Omega^{2}_{0}  $, and $\lambda_6 = 1.32475\Omega_{0} $,
with a total pulse area of $3.395\pi$. For this solution, the final phases are $\gamma_f = \varphi_f = 0.73\pi$, such that $\eta = 0$ and $\zeta = 0.73\pi$. The unitary directly generated by the pulse is therefore
\begin{equation}
U_{\mathrm{H};0.73\pi,0} =
\frac{1}{\sqrt{2}}
\begin{pmatrix}
1 & -e^{i 0.73\pi} \\
e^{-i 0.73\pi} & 1
\end{pmatrix}
\label{eq:U_H_type}
\end{equation}
%

\begin{figure} 
\centering
\includegraphics[width=1\columnwidth]{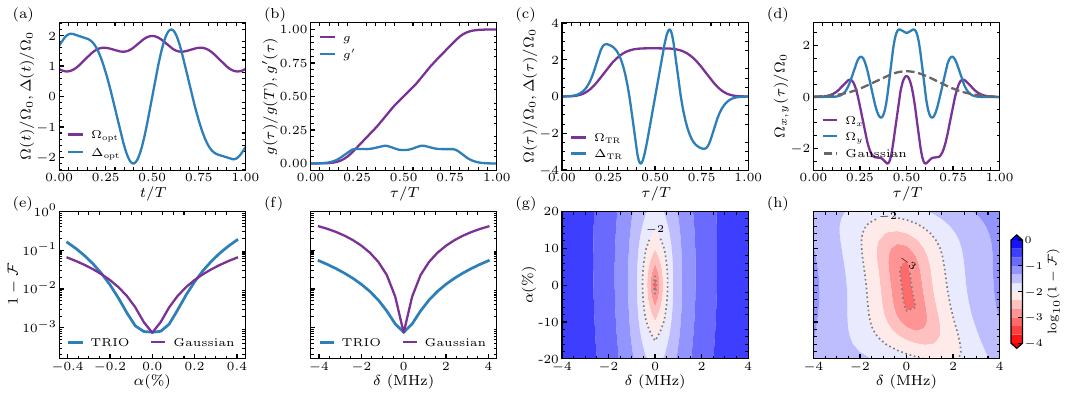}
\caption{\textbf{TRIO pulse for the H gate.}
\textbf{a}, The energy-optimal pulses ${\Omega_{\rm opt}, \Delta_{\rm opt}}$ for a two-level system. \textbf{b}, The time-rescaling function rescales the optimal pulse into 
\textbf{c}, a hardware-efficient pulse envelope and 
\textbf{d}, the corresponding experimentally implementable pulse shape. 
\textbf{e}, \textbf{f}, Robustness simulations: infidelity as a function of amplitude deviation $\alpha$ and detuning $\delta$, respectively, for TRIO (blue) and Gaussian (purple) pulses.
\textbf{g}, \textbf{h}, Infidelity as a function of the combined error for the Gaussian and the TRIO pulse, resepectively.
Note that \textbf{a}--\textbf{d} is the optimal result for a two-level system, while \textbf{e}--\textbf{h} consider the realistic hardware with parameters the same as those in \figref{fig:robust-three-level_derio_open}.
Note that the unit $\Omega_{0}  = \max[{\Omega_{\rm G}(t)}]$ refers to the maximum value of a Gaussian $\pi$-pulse, e.g., the dashed curve in \textbf{d}.
\label{fig:robust-two-level_hadmard}}
\end{figure}
 
The optimal solution is presented in \figpanel{fig:robust-two-level_hadmard}{a}. We next implement the TR technique with the low-order hyper-Gaussian pulse defined in \eqref{eq:hyper_gauss_appendix}. The time-rescaled pulse is illustrated in \figpanels{fig:robust-two-level_hadmard}{b}{c}.
Then, we transform the lab-frame controller into a phase-frame XY pulse as follows:
\beq
\Omega^{H;0.73\pi,0}_{\rm TRIO}(\tau) =
e^{-i\int_{0}^{\tau} \Delta^{H;0.73\pi,0}_{\rm TR}(s) ds}
\Omega^{H;0.73\pi,0}_{\rm TR}(\tau) ,
\eeq
as shown in \figpanel{fig:robust-two-level_hadmard}{d}. Since the TR preserves the time symmetry, the phase modulation 
$\phi_{\rm eff}(\tau_f) = \int_{0}^{\tau_f} \Delta^{H;0.73\pi,0}_{\rm TR}(s) ds < 10^{-4}$
in our simulation. In other words, the optimized $H$ gate requires no additional virtual-$Z$ correction to recover the target gate propagator under microwave phase modulation.

This operation has the balanced population transfer of a Hadamard-type gate, but differs from the conventional Hadamard operation
\begin{equation}
\label{eq:U_H}
H = \frac{1}{\sqrt{2}}
\begin{pmatrix}
1 & 1 \\
1 & -1
\end{pmatrix}
\end{equation}
by deterministic single-qubit phases. 
We compensate these phases with virtual-$Z$ rotations, defined as
$V_Z(\phi) = \mathrm{diag} (e^{-i \phi/2}, e^{i \phi/2})$. 
In general,
\begin{equation}
V_Z(\zeta) U_{\mathrm{H};\zeta,0} V_Z(\pi-\zeta) = -i H ,
\end{equation}
where the prefactor $-i$ is an irrelevant global phase. By combining \eqref{eq:U_H_type} and \eqref{eq:U_H}, we have
\begin{equation}
V_Z(0.73\pi) U_{\mathrm{H};0.730\pi,0} V_Z(0.27\pi) = -i H.
\end{equation}
Accordingly, the physical Hadamard-type pulse is preceded by a virtual-$Z$ rotation of $0.27\pi$ ($48.6^\circ$) and followed by a virtual-$Z$ rotation of $0.73\pi$ ($131.4^\circ$), yielding the standard Hadamard gate up to a global phase.

In particular, the $X_{\pi/2}$ gate, defined as
\begin{equation}
X_{\pi/2}: R_x(\pi/2) = \frac{1}{\sqrt{2}}
\begin{pmatrix}
1 & -i \\
-i & 1
\end{pmatrix},
\end{equation}
can be constructed from the SU(2) representative $-iH$ as
\begin{equation}
R_x(\pi/2)
= V_Z(-\pi/2) (-iH) V_Z(-\pi/2)
= V_Z(0.23\pi) U_{\mathrm{H};0.73\pi,0} V_Z(-0.23\pi) .
\label{eq:X2_vz}
\end{equation}
Similarly, the $Y_{-\pi/2}$ gate,
\begin{equation}
Y_{-\pi/2}: R_y(-\pi/2) = \frac{1}{\sqrt{2}}
\begin{pmatrix}
1 & 1 \\
-1 & 1
\end{pmatrix},
\end{equation}
is obtained as
\begin{equation}
R_y(-\pi/2)
= V_Z(-\pi) (-iH)
= V_Z(-0.27\pi) U_{\mathrm{H};0.73\pi,0} V_Z(0.27\pi) .
\label{eq:mY2_vz}
\end{equation}
Thus, no additional optimization of the physical pulse is required for the $X/2$ and $-Y/2$ gates: both are generated from the same Hadamard-type pulse by modifying only the virtual-$Z$ frame updates.
Table~\ref{tab:single_qubit_vz} summarizes the three single-qubit gates we implemented in a superconducting quantum processor and the phases we needed to adjust to do so.

\begin{table}
    \centering
    \begin{tabular}{l|c|c|c|c}
        \toprule
        \textbf{Gate} & \textbf{Physical pulse}
        & \textbf{$V_Z$ before} & \textbf{$V_Z$ after} & $\phi_{\rm eff}(\tau_f)$ \\
        \midrule
        $X$ & $U_{\mathrm{NOT}}$
        & $0$ & $0$ & $-0.537\pi$
        \\
        $X_{\pi/2}$ & $U_{\mathrm{H};0.730\pi,0}$
        & $-0.230\pi$ & $0.230\pi$ & 0
        \\
        $Y_{-\pi/2}$ & $U_{\mathrm{H};0.730\pi,0}$
        & $0.270\pi$ & $-0.270\pi$ & 0
        \\
        \bottomrule
    \end{tabular}
    \caption{Virtual-$Z$ corrections used to construct the single-qubit gates in this work.
    The ``before'' and ``after'' angles refer to the temporal order relative
    to the corresponding physical pulse.
    \label{tab:single_qubit_vz}}
\end{table}



\section{Experimental setup}

\begin{figure}
\centering
\includegraphics[width=0.7\columnwidth]{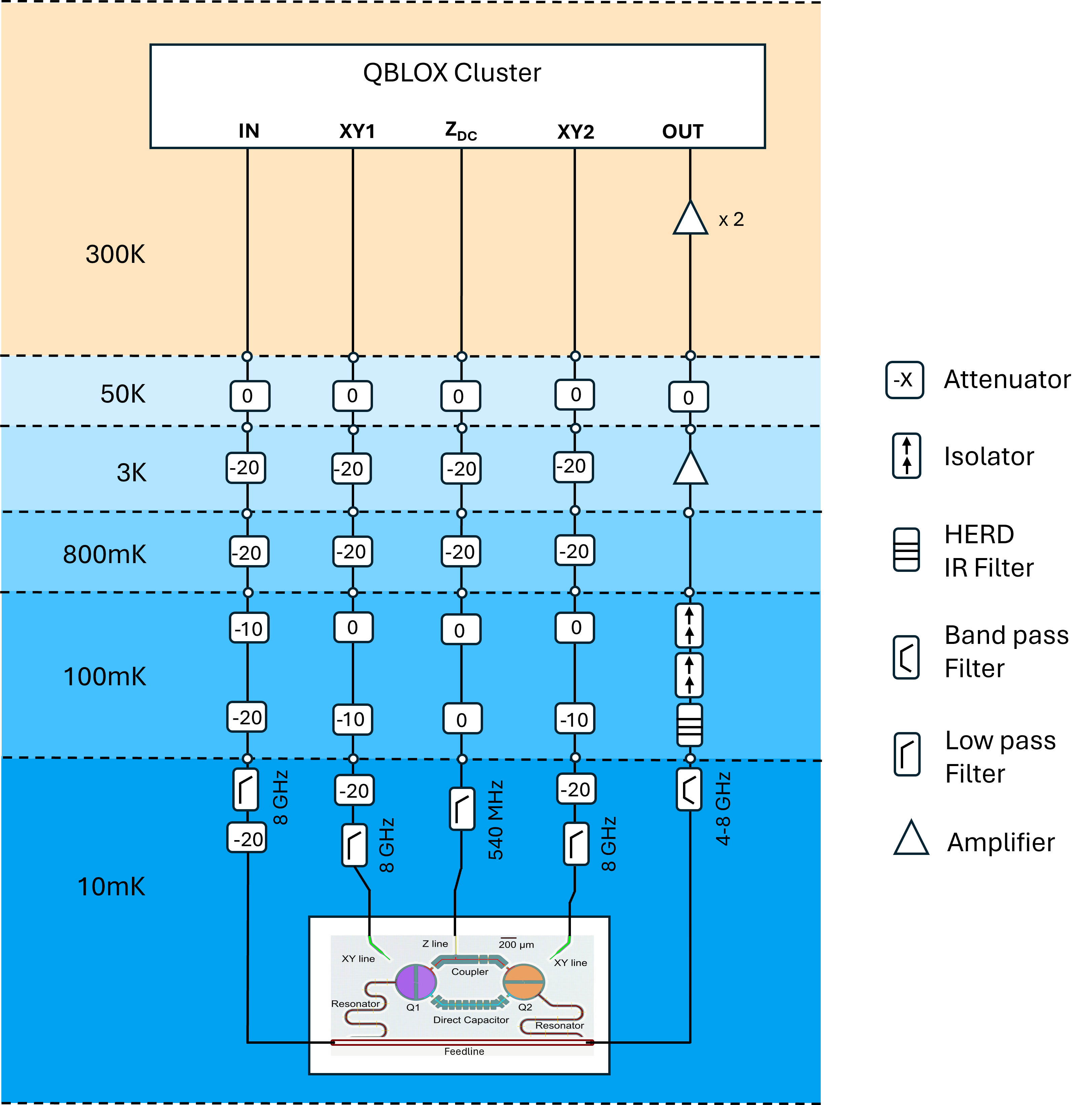}
\caption{\textbf{Experimental setup.}
The different temperature stages are color-coded according to temperature. The control hardware used is a Qblox Cluster. The XY$_1$ and XY$_2$ lines are the microwave drive lines for the qubits Q$_1$ and Q$_2$, respectively, and are used for single-qubit control. The $Z_{\mathrm{DC}}$ line provides the flux bias for the tunable coupler. The IN and OUT lines denote the readout input and output paths, respectively. Attenuation values are given in dB.}
\label{fig:wiring diagram}
\end{figure}

Figure~\ref{fig:wiring diagram} shows the full wiring diagram of the experimental setup. The device is mounted at the \SI{10}{\milli\kelvin} stage of a Bluefors LD250 dilution refrigerator. The microwave input lines pass through successive temperature stages, where they are strongly attenuated to suppress thermal photon noise. The readout return line is filtered using a band-pass filter and an HERD IR filter~\cite{rehammar2023low}, and isolated using directional isolators. The readout signal is first amplified by a high-electron-mobility transistor (HEMT) amplifier at \SI{4}{\kelvin} and then by two additional amplifiers at room temperature. Control signals and data acquisition at room temperature are provided by a Qblox Cluster equipped with QCM-RF, QRM-RF, and QCM modules, operating at a \SI{1}{\giga\hertz} sampling rate. Single-qubit control is implemented using microwave pulses generated by the QCM-RF modules and applied through the XY drive lines. Readout signals are sent to the device through the IN line, while the output signal from the device is digitized through the OUT line using the QRM-RF module. The DC flux bias applied to the tunable coupler is supplied by the QCM module through the $Z_{\mathrm{DC}}$ line.

\begin{figure}
\centering
\includegraphics[width=0.6\columnwidth]{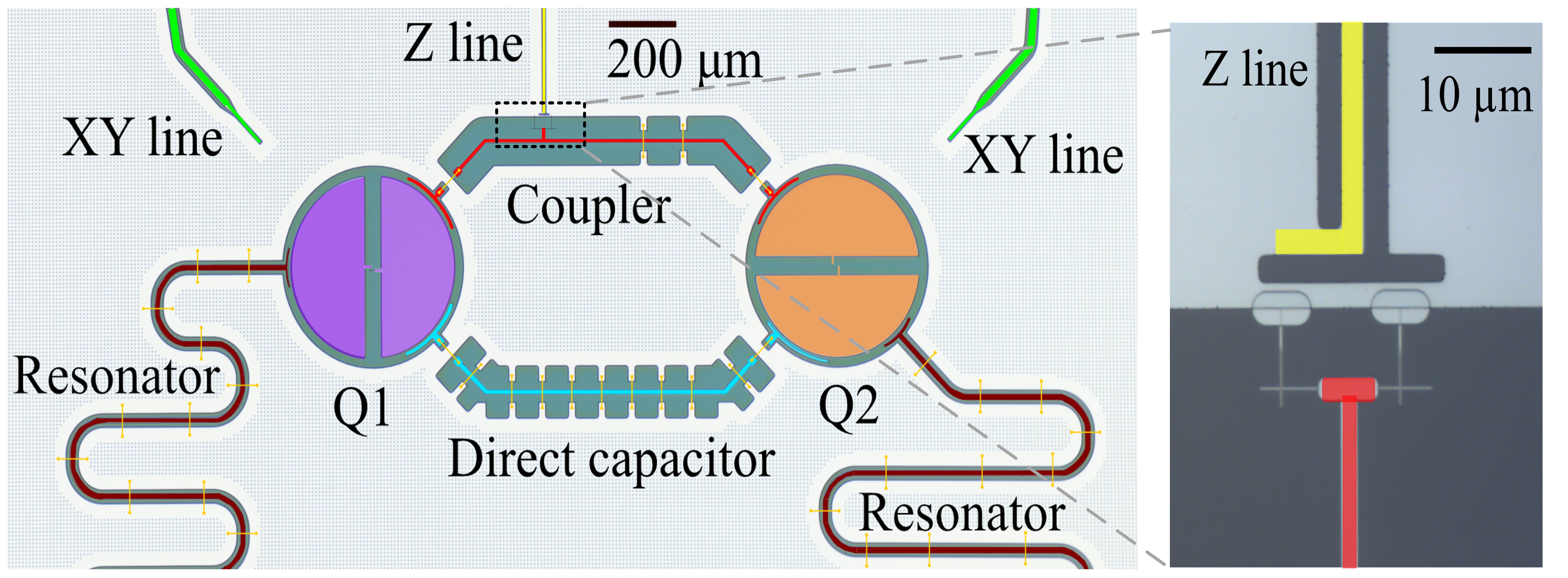}
\caption{\textbf{False-colored micrograph of the device.}
The device consists of two fixed-frequency transmon qubits and one flux-tunable coupler. In the experiments, Q$_1$ is used as the measured qubit. The coupler is flux-tuned, and its interaction with Q$_1$ shifts the Q$_1$ frequency, producing the detuning $\delta$. The right-hand image shows a zoomed-in view of the Z-line (yellow), which is shorted to ground and inductively coupled to the SQUID loop of the coupler.} 
\label{fig:2qb_device}
\end{figure}

The device, shown in more detail in \figref{fig:2qb_device}, consists of two fixed-frequency transmon qubits, Q$_1$ and Q$_2$, coupled through a flux-tunable transmon coupler. Each qubit is connected to a dedicated readout resonator. In this work, the gate-fidelity measurements are performed on Q$_1$, whose frequency, coherence times, and other parameters are summarized in \tabref{tab:device_parameters}. The device is fabricated on a high-resistivity silicon substrate using aluminum thin-film processing, with the Josephson junctions defined by double-angle evaporation, following the fabrication process described in Refs.~\cite{aggarwal2025mitigating, biznarova2024mitigation, chayanun2024characterization}.

\begin{table}
    \centering
    \caption{Measured qubit Q$_1$ parameters, coherence properties, and readout fidelity.}
    \label{tab:device_parameters}
    \begin{tabular}{lcc}
        \toprule
        \textbf{Parameter} & \textbf{Q$_1$}  \\
        \midrule
        Qubit frequency $(\omega_0/2\pi)$
            & \SI{4.7744}{\giga\hertz} \\
        Qubit anharmonicity $(\xi/2\pi)$
            & \SI{239}{\mega\hertz} \\
        Resonator frequency $(\omega_r/2\pi)$
            & \SI{6.997}{\giga\hertz}  \\
        Dispersive shift $(\chi_r/2\pi)$
            & \SI{0.420}{\mega\hertz} \\
        Relaxation time $(T_1)$
            & \SI{75}{\micro\second} \\
        Decoherence time $(T_2^*)$
            & \SI{55}{\micro\second}  \\
        Three-state assignment fidelity
            & \SI{91.5}{\percent} \\
        $(F_{\mathrm{Assign}(\ket{0}, \ket{1}, \ket{2}})$
            & & \\
        \bottomrule
    \end{tabular}
\end{table}
 

\subsection{Qubit characterization and calibration}
\label{sec:qubit_calibration}

The experiments are performed on a fixed-frequency transmon coupled to a flux-tunable coupler. We first determine the readout-resonator frequency and the qubit transition frequency by spectroscopy. The qubit drive frequency is subsequently refined using Ramsey measurements with progressively longer evolution times. The microwave amplitude is calibrated using Rabi oscillations and further optimized by applying repeated nominal $\pi$ rotations, which amplify small pulse-area errors. The calibrated Gaussian pulses provide the reference operations for state preparation, tomographic rotations, and randomized benchmarking.

\begin{figure} 
\centering
\includegraphics[width=\columnwidth]{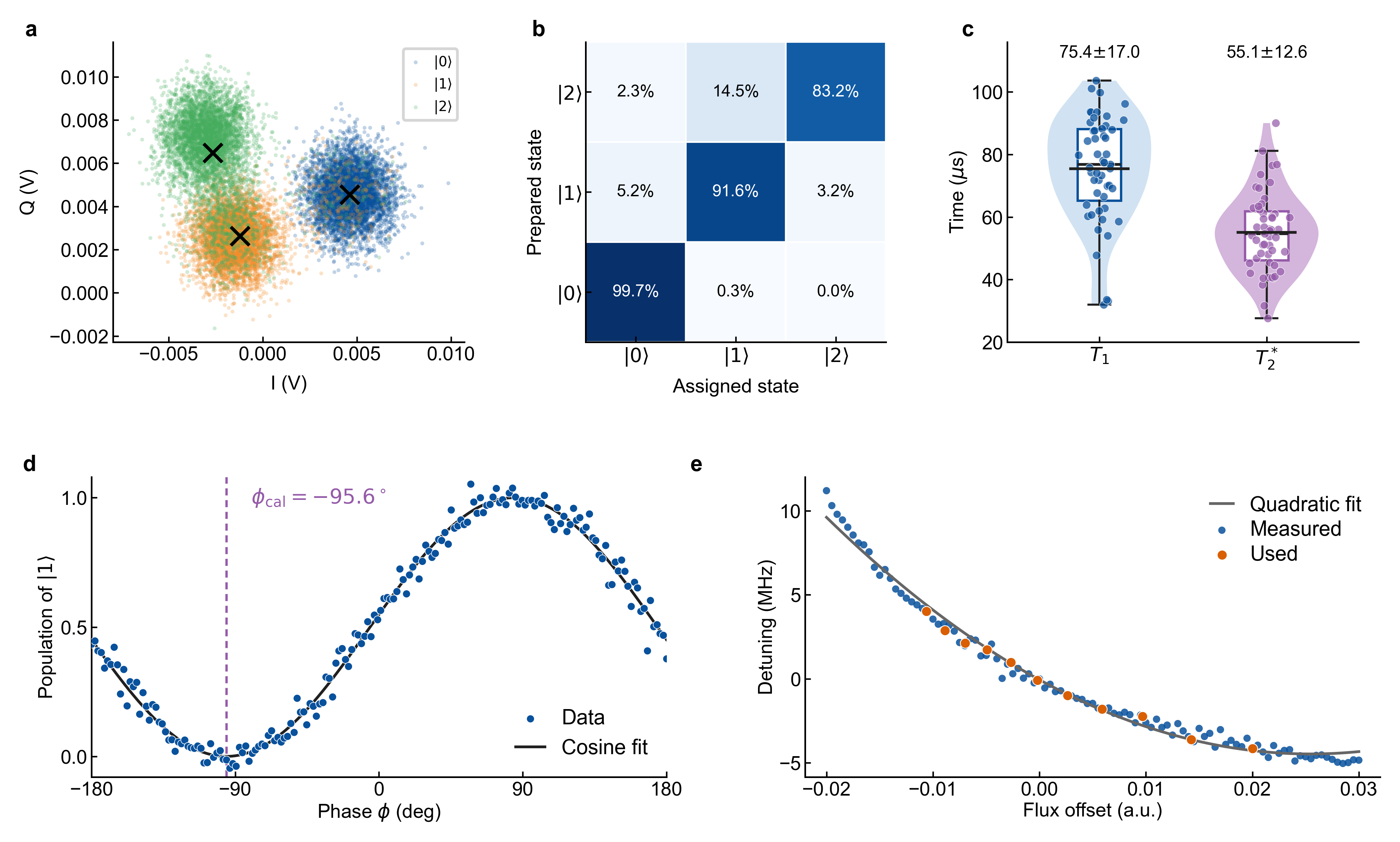}
\caption{
\textbf{Qubit characterization and gate calibration.}
\textbf{a}, Single-shot IQ distributions for the prepared states $\ket{0}$, $\ket{1}$, and $\ket{2}$, with $10^4$ measurements per state. Black crosses mark the centroids used for nearest-centroid state assignment.
\textbf{b}, Corresponding assignment probabilities, including both state-preparation and readout errors.
\textbf{c}, Distributions of relaxation times $T_1$ and Ramsey coherence times $T_2^*$ obtained from repeated measurements, with mean values of \SI{75.4}{\micro\second} and \SI{55.1}{\micro\second}, respectively. Points represent individual measurements.
\textbf{d}, Phase calibration of the TRIO $X$ gate using a Ramsey-type sequence. Points denote measured excited-state populations. The solid curve is a cosine fit. The dashed line marks the fitted minimum at $\phi_{\mathrm{cal}} = -95.6^\circ$.
\textbf{e}, Qubit-frequency detuning as a function of the voltage offset applied to the coupler flux line. Blue points denote measurements, the grey curve is a quadratic fit, and orange points indicate the operating points used in the robustness measurements.
}
\label{fig:exp_calibration}
\end{figure}

To resolve population outside the computational subspace (spanned by the states $\ket{0}$ and $\ket{1}$), we calibrate dispersive readout for the three lowest transmon levels.
The states $\ket{0}$, $\ket{1}$, and $\ket{2}$ are prepared by initialization, a calibrated $X_{01}$ pulse, and consecutive $X_{01}$ and $X_{12}$ pulses, respectively. The amplitude of the $X_{12}$ pulse is calibrated independently using Rabi oscillations on the $\ket{1} \leftrightarrow \ket{2}$ transition.
For each prepared state, we acquire $10^4$ single-shot measurements of the integrated complex readout signal $z = I + iQ$.
The resulting IQ distributions and their centroids are shown in \figpanel{fig:exp_calibration}{a}.

Each measurement outcome is assigned to the nearest calibration centroid in the IQ plane,
\begin{equation}
\hat{j}(z) = \underset{j \in \{0, 1, 2\}}{\operatorname{argmin}} \mleft| z - \bar{z}_j \mright|^2,
\end{equation}
where $\bar{z}_j$ is the mean signal for the preparation of state $\ket{j}$. The assignment matrix is defined as
$M_{ij} = P(\hat{j} = j\,|\,\mathrm{preparation\ of}\ \ket{i})$. 
For the calibration shown in \figpanel{fig:exp_calibration}{b}, the diagonal assignment probabilities are \SI{99.7}{\percent}, \SI{91.6}{\percent}, and \SI{83.2}{\percent} for $\ket{0}$, $\ket{1}$, and $\ket{2}$, respectively. These probabilities characterize the combined state-preparation and readout procedure.

The TRIO waveform is sampled at \SI{1}{\nano\second} intervals and scaled relative to the calibrated Gaussian $\pi$ pulse using the pulse area prescribed by the control design. We then calibrate its DRAG correction and residual phase~\cite{DRAG}.
The DRAG coefficient, i.e., $-\beta/\xi$ in \eqref{eq:D_drag_complex_envelope}, is scanned while applying even numbers $N_X$ of phase-compensated TRIO $X$ pulses. Ideally, each such sequence returns the qubit to its initial state.
Repeated application amplifies residual control errors. 
The coefficient is selected by minimizing the readout signal averaged over the sampled sequence lengths. 

Energy relaxation and Ramsey coherence are characterized by varying the delay following an excitation pulse and between two $\pi/2$ pulses, respectively. 
Figure~\figpanelNoPrefix{fig:exp_calibration}{c} summarizes repeated measurements, with characteristic relaxation and Ramsey coherence times of \SI{75.4}{\micro\second} and \SI{55.1}{\micro\second}, respectively.
 
The phase correction is determined using a Ramsey-type sequence.
A calibrated $Y_{\pi/2}$ rotation prepares an equatorial state, followed by the TRIO pulse and a $Y_{-\pi/2}$ analysis rotation.
Experimentally, we perform a phase-sweep sequence:
\beq
Y_{\pi/2} \rightarrow U_{\mathrm{p}} \rightarrow R_{z}(\theta) \rightarrow Y_{-\pi/2} \rightarrow \text{Measurement}, 
\eeq
with a phase reset at the beginning of each shot to avoid inter-shot frame accumulation. For an ideal $U_{\mathrm{p}} = X R_{z}(\phi)$, this yields an excited-state probability
\begin{equation}
P_{1}(\theta) = \frac{1 - \cos(\phi + \theta)}{2}.
\end{equation}
The phase at the fitted minimum defines the compensation applied through a virtual-$Z$ operation. The calibration in \figpanel{fig:exp_calibration}{d} yields 
$\phi_{\mathrm{cal}} \simeq -95.6^\circ (-0.531\pi)$,
which is close to the theoretical prediction $-0.537\pi$.
The phase is recalibrated after optimization of the DRAG coefficient.

For the Hadamard gate, the TRIO pulse is supplemented with virtual-$Z$ corrections before and after the microwave pulse,
\begin{equation}
H_{\mathrm{TRIO}} = V_Z(\pi-\theta_{\mathrm{pre}}) U_{\mathrm{TRIO}} V_Z(\theta_{\mathrm{pre}}) ,
\end{equation}
where operators act from right to left. The two correction phases are determined using complementary calibration sequences that probe the target transformations $H \ket{0} = \ket{+}$ and $H \ket{+} = \ket{0}$, with $\ket{+} = (\ket{0} + \ket{1} / \sqrt{2}$.

To calibrate $\theta_{\mathrm{pre}}$, the qubit is initialized in $\ket{0}$, and the TRIO pulse is followed by the variable virtual-$Z$ correction and a calibrated $Y_{-\pi/2}$ analysis pulse. The correction phase is swept over a full $2\pi$ interval, and the resulting readout signal is fitted to
$S(\theta) = a + b \cos\theta + c \sin\theta$.
The phase corresponding to the ground-state return condition is selected.
To calibrate $\theta_{\mathrm{pre}}$, a calibrated $Y_{\pi/2}$ pulse first prepares $\ket{+}$. The variable virtual-$Z$ correction is then applied before the TRIO pulse, followed by measurement in the computational basis.
The same sinusoidal fitting procedure determines the phase that returns the qubit to $\ket{0}$.
Together, these complementary measurements determine the input and output phase corrections required for the Hadamard operation.
The calibrated corrections are implemented as updates to the microwave reference phase, without additional physical pulses.

Controlled frequency offsets are introduced through the coupler flux line. We measure the qubit transition frequency as a function of the applied flux-control offset $v$ and fit the local response to a quadratic function,
\begin{equation}
\Delta f(v) \equiv f_{01}(v) - f_{01}(0) = a_2 v^2 + a_1 v + a_0 .
\end{equation}
The fitted relation is inverted within the calibrated range to determine the offsets used in the robustness measurements [\figpanel{fig:exp_calibration}{e}].
Here, $\Delta f$ is expressed in frequency units; the corresponding angular-frequency detuning is $2\pi\Delta f$.


\subsection{Interleaved randomized benchmarking}
\label{sec:irb_leakage}

We characterize the implemented gates using interleaved randomized benchmarking (IRB)~\cite{Magesan2012IRB}.
A reference sequence consists of $m$ randomly sampled single-qubit Clifford operations followed by a recovery Clifford that ideally returns the qubit to its initial state.
The Clifford operations are compiled into calibrated rotations about the $x$ and $y$ axes.
For the interleaved sequence, the gate under test, including its calibrated virtual-$Z$ correction, is inserted after each random Clifford. 
The recovery operation is calculated from the complete ideal sequence, including the interleaved gates.
Reference and interleaved sequences use the same random seeds.

For the fidelity measurements, the averaged complex readout signal is projected onto the axis defined by the $\ket{0}$ and $\ket{1}$ calibration signals.
The resulting reference and interleaved survival probabilities are fitted independently to
\begin{equation}
P_\nu(m) = A_\nu p_\nu^m + B_\nu , \qquad
\nu \in \{\mathrm{ref}, \mathrm{int}\},
\end{equation}
where $A_\nu$ and $B_\nu$ absorb sequence-independent preparation and measurement offsets, and $p_\nu$ is the fitted decay parameter.
The reference Clifford fidelity and the IRB estimate of the target-gate fidelity are~\cite{Magesan2012IRB}
\begin{align}
F_{\mathrm{Cl}} &= 1 - \frac{1 - p_{\mathrm{ref}}}{2} , \\
F_{\mathrm{IRB}} &= 1 - \frac{1}{2} \mleft( 1 - \frac{p_{\mathrm{int}}}{p_{\mathrm{ref}}} \mright) .
\end{align}
These expressions use the standard single-qubit IRB approximation. Fit-parameter uncertainties are propagated through the fidelity estimator.

For robustness measurements, the complex envelope of the interleaved pulse is multiplied by $(1 + \alpha)$ to introduce a relative amplitude error $\alpha$. A calibrated flux pulse is applied during the interleaved operation to introduce a qubit-frequency offset. In this schedule, the flux pulse begins \SI{4}{\nano\second} before the microwave pulse and ends \SI{4}{\nano\second} after it. The reference Clifford operations retain their nominal calibrations. The extracted fidelity therefore characterizes the interleaved operation under the imposed error conditions.

Leakage is characterized using the three-state single-shot readout with IRB. For each sequence, we determine the fraction of outcomes assigned to $\ket{2}$ using the IQ classifier described in \secref{sec:qubit_calibration}. The assigned leakage populations are averaged over random sequences; their statistical uncertainties are estimated from the standard error across sequences.

We fit the reference and interleaved leakage traces to a two-subspace rate model~\cite{Wood2018Leakage},
\begin{equation}
p_{2,\nu}(m) = p_{2,\nu}^{\infty} + \mleft[ p_{2,\nu}(0) - p_{2,\nu}^{\infty} \mright] (1 - L_\nu - S_\nu)^m,
\end{equation}
with
\begin{equation}
p_{2,\nu}^{\infty} = \frac{L_\nu}{L_\nu + S_\nu} .
\end{equation}
Here, $L_\nu$ and $S_\nu$ describe effective transfer out of and back into the computational subspace per sequence step, respectively, and $p_{2,\nu}(0)$ is a fitted offset. One reference step contains a Clifford operation, whereas one interleaved step contains a Clifford and the target gate.

The analysis estimates the additional leakage associated with the interleaved gate as
\begin{equation}
L_G^{\mathrm{est}} = 1 - \frac{1 - L_{\mathrm{int}}}{1 - L_{\mathrm{ref}}} \simeq L_{\mathrm{int}} - L_{\mathrm{ref}} .
\end{equation}
The first expression neglects return from the leakage subspace within a composed reference--target step; the second is its small-error limit. The difference between the initial slopes of the two leakage traces provides an additional diagnostic.
In the present analysis, the traces are fitted directly to the assigned $\ket{2}$ fractions without inversion of the readout assignment matrix. The extracted quantities are therefore effective leakage estimates that retain sensitivity to faulty readout assignment.

\begin{figure*}
\centering
\includegraphics[width=1\columnwidth]{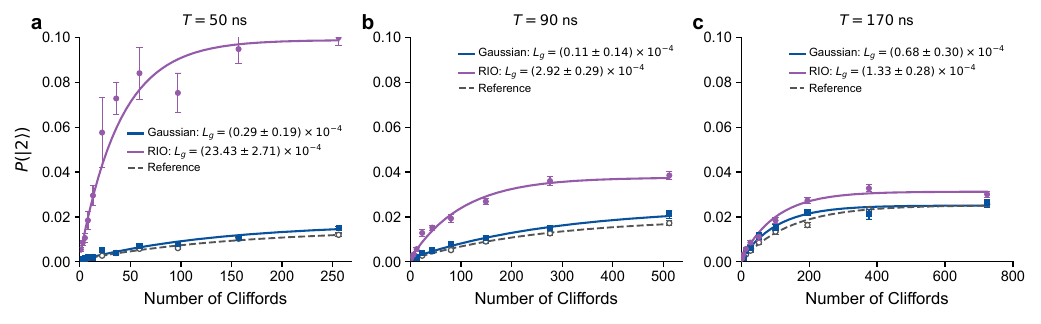}
\caption{\textbf{Leakage measured by interleaved randomized benchmarking.}
Panels \textbf{a--c} correspond to gate durations $T = 50$, 90, and \SI{170}{\nano\second}, respectively.
The leakage-state population $P(\ket{2})$ is plotted as a function of the number of Cliffords.
Blue squares and purple circles denote Gaussian and RIO interleaved sequences, respectively; open grey circles denote the reference data.
Solid and dashed lines show fits to the interleaved and reference data, respectively.
The legends report the gate leakage estimates $L_g$.
Error bars represent the standard error of the mean; quoted uncertainties are propagated $1\sigma$ fit uncertainties.
}
\label{fig:irb_leakage}
\end{figure*}

Figure~\ref{fig:irb_leakage} shows that RIO exhibits substantially enhanced leakage at short gate durations, accompanied by a reduction in interleaved gate fidelity. At gate duration $T = \SI{50}{\nano\second}$, the leakage of RIO ($\sim 2.3\times10^{-3}$) is markedly higher than that of the Gaussian gate ($\sim 2.9\times 10^{-5}$), indicating that population transfer outside the computational subspace is an important contributor to its reduced fidelity. Increasing the RIO gate duration suppresses leakage and improves fidelity, consistent with leakage being a significant limitation in the short-duration regime.


\subsection{Quantum process tomography}
\label{sec:qpt}

We perform quantum process tomography (QPT) of the experimentally implemented $X$ gate. Here, we use the Choi-matrix representation ${J}_{\mathcal{E}}$ to represent the underlying quantum channel $\mathcal{E}$ as
\begin{equation}
{J}_{\mathcal{E}} =
\sum_{i,j} \ket{i}\bra{j}_{\mathcal{H}_{\rm in}} \otimes
\mathcal{E} \mleft( \ket{i}\bra{j} \mright)_{\mathcal{H}_{\rm out}} ,
\end{equation}
where $\mathcal{H}_{\rm in}$ and $\mathcal{H}_{\rm out}$ denote the system input and output Hilbert spaces, respectively. In the Choi-matrix representation, the action of the underlying channel ${\mathcal{E}}$ on the given input state $\varrho$ is given by
\begin{equation}
{\mathcal{E}} (\varrho) = \text{Tr}_{\mathcal{H}_{\rm in}}
\mleft[ \mleft(\varrho^{\mathrm{T}} \otimes \mathbb{I}_{\mathcal{H}_{\rm out}} \mright) {J}_{\mathcal{E}} \mright] ,
\label{eq2:choi_matrix}
\end{equation}
where $\text{Tr}_{\mathcal{H}_{\rm in}}$ denotes the partial trace over the input space, $\varrho^{\mathrm{T}}$ is the transpose of the input state $\varrho$, and $\mathbb{I}_{\mathcal{H}_{\rm out}}$ is the identity operator on the output space.

To characterize the experimentally implemented $X$ gate, we first prepare informationally complete probe states $\varrho \in \{ \ket{0}, \ket{1}, \ket{+}, \ket{+i} \}$. Then we apply the gate under test, followed by measurements in different bases $\Pi \in \mleft\{ U^\dag \Pi_{\pm z} U \; \vert \; U \in \{ I, R_x,R_y \} \mright\}$, resulting in 12 probe circuits in total. From each probe circuit, specified by the input state $\varrho_i$ and the measurement operator $\Pi_j$, we collect a data point $d_{ij}$, resulting in 24 data points. We then form a constrained optimization problem as
\begin{equation}
\min_{ {J}_{\mathcal{E}}} \sum_{i,j} \mleft[ d_{ij} - \text{Tr}[\Pi_j \mathcal{E}(\varrho_i)] \mright]^2 , \quad \text{s.t.} \quad  
{J}_{\xi} \succeq 0 \; \text{(CP)} \;\;\; \text{and} \;\;\; 
\text{Tr}_{\mathcal{H}_{\rm out}}({J}_{\xi}) = \mathbb{I}_{\mathcal{H}_{\rm in}} \; \text{(TP)} ,
\label{eq:projgd}
\end{equation}
where $\mathcal{E}(\varrho_i)$ represents the experimental output state produced after applying the target gate to the initial state $\varrho_i$ [see \eqref{eq2:choi_matrix}]. Here, we employ a projected gradient-descent algorithm~\cite{gaikwad-qst-2025} to solve the QPT optimization problem described in \eqref{eq:projgd} with appropriate parameterization of the Choi matrix ${J}_{\mathcal{E}}$ that makes it respect the constraints of being completely positive (CP) and trace-preserving (TP), as required for a valid quantum operation~\cite{julia-prr-2025}. Finally, to obtain the average process fidelity $\langle \mathcal{F}\rangle$ and the corresponding standard deviation $\Delta \mathcal{F}$, we repeated the full QPT procedure 10 times, resulting in $\mathcal{F} = 0.9993 \pm 0.0005$.

\end{document}